\documentclass{aastex}
\usepackage{amsmath}
\usepackage{multirow}
\usepackage{array}
\usepackage{float}
\usepackage[caption = false]{subfig}
\usepackage{graphicx}
\usepackage[backref,breaklinks,colorlinks,citecolor=blue]{hyperref} 
\usepackage[all]{hypcap} 

\usepackage{cleveref}
\usepackage[section]{placeins}	
\usepackage{natbib}
\newcommand{\minus}{\scalebox{0.55}[1.0]{$-$}}
\begin{document}
\title{The BRITE Constellation nanosatellite mission:
Testing, commissioning and operations\thanks{Based
on data collected by the BRITE Constellation satellite mission, designed, built,
launched, operated and supported by the Austrian Research Promotion Agency (FFG),
the University of Vienna, the Technical University of Graz, the Canadian Space
Agency (CSA), the University of Toronto Institute for Aerospace Studies (UTIAS), the
Foundation for Polish Science \& Technology (FNiTP MNiSW), and National Science
Centre (NCN).}}
\author{H.~Pablo\altaffilmark{1}, G.~N.~Whittaker\altaffilmark{2}, 
A.~Popowicz\altaffilmark{3}, S.~M.~Mochnacki\altaffilmark{2}, R.~Kuschnig\altaffilmark{4,8}, C.~C.~Grant\altaffilmark{5}, A.~F.~J.~Moffat\altaffilmark{1}, S.~M.~Rucinski\altaffilmark{2},
J.~M.~Matthews\altaffilmark{6},
A.~Schwarzenberg-Czerny\altaffilmark{7}, G.~Handler\altaffilmark{7}, W.~W.~Weiss\altaffilmark{8}, 
D.~Baade\altaffilmark{9},
G.~A.~Wade\altaffilmark{10},
E.~Zoc{\l}o\'nska\altaffilmark{7}, T.~Ramiaramanantsoa\altaffilmark{1},
M.~Unterberger\altaffilmark{4},
K.~Zwintz\altaffilmark{11}, 
A.~Pigulski\altaffilmark{12}
J.~Rowe\altaffilmark{13}, 
O.~Koudelka\altaffilmark{4}, P.~Orlea\'nski\altaffilmark{14},  A.~Pamyatnykh\altaffilmark{7},
C.~Neiner\altaffilmark{15},
R.~Wawrzaszek\altaffilmark{14},  G.~Marciniszyn\altaffilmark{7}, P.~Romano\altaffilmark{4}, G.~Wo\'zniak\altaffilmark{7},
T.~Zawistowski\altaffilmark{14},
R.~E.~Zee\altaffilmark{5}}

\altaffiltext{1}{D\'epartement de physique, Universit\'e de Montr\'eal, C.P.~6128, Succursale center-Ville, Montr\'eal, Qu\'ebec, H3C\,3J7, Canada,\\ et center de recherche en astrophysique du Qu\'ebec (CRAQ); hpablo@astro.umontreal.ca}
\altaffiltext{2}{Department of Astronomy \& Astrophysics, University of Toronto, 50 St.~George Street, Toronto, Ontario, M5S\,3H4, Canada}
\altaffiltext{3}{Silesian University of Technology, Institute of Automatic Control, Gliwice, Akademicka 16, Poland}
\altaffiltext{4}{Graz University of Technology, Institute of Communication Networks and Satellite Communications, Inffeldgasse 12, 8010 Graz, Austria}
\altaffiltext{5}{Space Flight Laboratory, University of Toronto, 4925 Dufferin St., Toronto, Ontario, M3H5T6, Canada}
\altaffiltext{6}{Dept.~of Physics \& Astronomy, The University of British Columbia, 6224 Agricultural Road, Vancouver, B.C., V6T\,1Z1, Canada}
\altaffiltext{7}{Centrum Astronomiczne im.~M.\,Kopernika, Polska Akademia Nauk, Bartycka 18, 00-716 Warszawa, Poland}
\altaffiltext{8}{Institut f\"ur Astrophysik, Universit\"at Wien, T\"urkenschanzstrasse 17, 1180 Wien, Austria}
\altaffiltext{9}{European Organisation for Astronomical Research in the Southern Hemisphere (ESO), Karl-Schwarzschild-Str. 2, 85748 Garching, Germany}
\altaffiltext{10}{Department of Physics, Royal Military College of Canada, PO Box 17000, Station Forces, Kingston, Ontario, K7K\,7B4, Canada}
\altaffiltext{11}{Institut f\"ur Astro- und Teilchenphysik, Universit\"at Innsbruck, Technikerstrasse 25/8, 6020 Innsbruck, Austria}
\altaffiltext{12}{Instytut Astronomiczny, Uniwersytet Wroc{\l}awski, Kopernika 11, 51-622 Wroc{\l}aw, Poland}
\altaffiltext{13}{Institut de recherche sur les exoplan\`etes, iREx, D\'epartement de physique, Universit\'e de Montr\'eal, Montr\'eal, QC, H3C 3J7, Canada}
\altaffiltext{14}{Centrum Bada\'n Kosmicznych, Polska Akademia Nauk, Bartycka 18A, 00-716 Warszawa, Poland}
\altaffiltext{15}{LESIA, Observatoire de Paris, PSL Research University, CNRS, Sorbonne Universit\'{e}s, UPMC Univ. Paris 06, Univ. Paris Diderot, Sorbonne Paris Cit\'{e}, 5 place Jules Janssen, 92195 Meudon, France}


\begin{abstract}
BRITE (BRIght Target Explorer) Constellation, the first nanosatellite mission applied to astrophysical research, is a collaboration among Austria, Canada and Poland. The fleet of satellites (6 launched; 5 functioning) performs precise optical photometry of the brightest stars in the night sky. A pioneering mission like BRITE - with optics and instruments restricted to small volume, mass and power in several nanosatellites, whose measurements must be coordinated in orbit - poses many unique challenges.  We discuss the technical issues, including problems encountered during on-orbit commissioning (especially higher-than-expected sensitivity of the CCDs to particle radiation). We describe in detail how the BRITE team has mitigated these problems, and provide a complete overview of mission operations. This paper serves as a template for how to effectively plan, build and operate future low-cost niche-driven space astronomy missions.
\end{abstract}


\section{Introduction}
\label{sect:intro}

Photometric variability is a key diagnostic in stellar astrophysics, from rotational modulation and activity changes to pulsations which enable us to probe internal structure and age via asteroseismology \citep{aerts2010}.  To most fully measure and interpret those variations, continuous coverage over long time scales is essential.  This is difficult to accomplish from the ground, even with global observatory networks like the Whole Earth Telescope \citep{wet}.  The combined need for long ungapped time coverage and high photometric precision, free of atmospheric effects, prompted the push to space-based missions in the last decade or so.

This push began with \textit{MOST} \citep{most2, most} and was followed soon after by CoRoT \citep{auvergne09} and Kepler \citep{kep1, kep2}. These missions made incredible strides in stellar and exoplanetary sciences, but all had their own limitations.  MOST could monitor much of the sky, but only a small number of stars at a time in a given small target field, and even smaller number of those could be monitored continuously. CoRoT and Kepler monitored much larger numbers of stars, but in small specific patches of the sky. Their target fields contained only small numbers of stars brighter than V $\approx$ 6. K2 while having access to a larger portion of the sky and many more bright targets, has issues with brighter stars.  None of these missions operated with effectively more than one filter, though CoRoT did make an attempt.  There was still a niche for a mission that offered high photometric precision, full sky coverage and multi-filter capability, for a sample of the brightest stars, which tend to be the most intrinsically luminous (i.e., massive and/or highly evolved).

BRITE Constellation extends the parameter space of space photometry missions, with nearly all-sky coverage in two wavelength ranges of hundreds of the most luminous stars in the Galaxy - all at relatively low cost \citep{paper1} .  Three partner nations (Austria, Canada and Poland) each contributed a pair of nanosatellites (mass 7 kg; 3-axis-stablized). The BRITE network is designed to collect optical photometry of millimagnitude precision (Popowicz et al. 2016, in prep; hereafter Paper III) in light curves of high cadence (20 - 25 s between consecutive exposures) and long duration (up to 6 months) through red and blue filters. The features of the six BRITE nanosatellites are listed in Table 1; only five are currently operating in orbit. The Austrian satellites are UniBRITE (UBr) and BRITE-Austria (BAb), the Polish are BRITE-Lem (BLb) and BRITE-Heweliusz (BHr), and the Canadian are BRITE-Toronto (BTr) and BRITE-Montr\'eal (BMb, which did not deploy correctly into orbit); where r and b refer to the satellites equipped with red and blue filters, respectively. 

This is Paper II in a series of publications that address the technical aspects of the BRITE mission.\hspace{1mm}
 The first paper in the series, \cite{paper1}, shall hereafter be referred to as Paper I. This paper provides a comprehensive history of the development of BRITE, the overall design of each satellite, and an explanation of the objectives that have been the driving forces behind the mission. Paper III in the series will be a description of the BRITE data reduction pipeline.

BRITE's prime directive is to observe bright stars ($V\le$\,4~mag), and shed light on their internal and surface dynamics. Among the benefits that BRITE offers are:
\begin{itemize}
\item{\textbf{A test bed for future astronomical surveys with small satellites.} The combination of cutting-edge science with small low-cost instruments in space has come about thanks to the advent of miniature reaction wheels \citep{sinclair07}, and miniature star trackers that can achieve arc-minute pointing accuracy \citep{enright10} for the low inertias of nanosats. This mission can provide valuable lessons for future endeavors into space-based astronomy, which is now becoming more accessible to universities and modest astronomical institutions. The BRITE nanosatellites are extremely cost effective with one pair costing less than one percent of Kepler's  $\approx \$600\,\rm{million}$ price tag \citep{kepcost}, albeit with many more limitations, providing the opportunity for niche-based space astronomy. Students can gain first-hand experience of space missions, from the design and building phases, through to launches and subsequent data reduction, analysis and interpretation.}
\item{\textbf{Bright stars are optimal astrophysical laboratories.} Bright stars can be studied in many ways with relatively high signal-to-noise, e.g spectroscopy at various wavelengths such as optical, UV and infrared. The BRITE mission has embraced observations to support the data collected by its satellites through the establishment of a Ground-Based Observation Team (GBOT), described in Section ~\ref{sect:gbot}.}

\end{itemize}
This paper describes (a) the detectors in Section 2 and key results from their  pre-flight testing and evaluation in Section 3; (b) the on-orbit commissioning (Section 4) and subsequent operations of the BRITE satellites (Section 5); and (c) the effects of radiation on the BRITE CCDs along with (d) the strategies that have been applied to compensate for those effects (Section 6). 


\section{The BRITE Instruments}
\label{sect:design}

The BRITE spacecraft designs are based on the Generic Nanosatellite Bus developed by the University of Toronto Institute for Aerospace Studies - Space Flight Laboratory (SFL) for their CanX\footnote{CanX is one of many programs run at SFL, see \url{http://utias-sfl.net} for more details.} (Canadian Advanced Nanospace eXperiment) program. The CanX nanosatellites were the first to include compact star-trackers, which paved the path for precise space-based photometry with the BRITE nanosatellites.
A comprehensive description of the satellite design is provided in Paper I, which includes details of the Attitude Determination and Control System (ADCS), telescope, instrument onboard computer (IOBC), power budget, and communications. To provide context for the descriptions of commissioning and operations later in this paper, we give brief descriptions of the hardware below, including additional information on the optics, focusing, CCDs, and thermal control systems.

\begin{figure}[htbp]
\begin{center}
\includegraphics[trim=0cm 0cm 0cm 0cm,clip,width=0.4\textwidth]{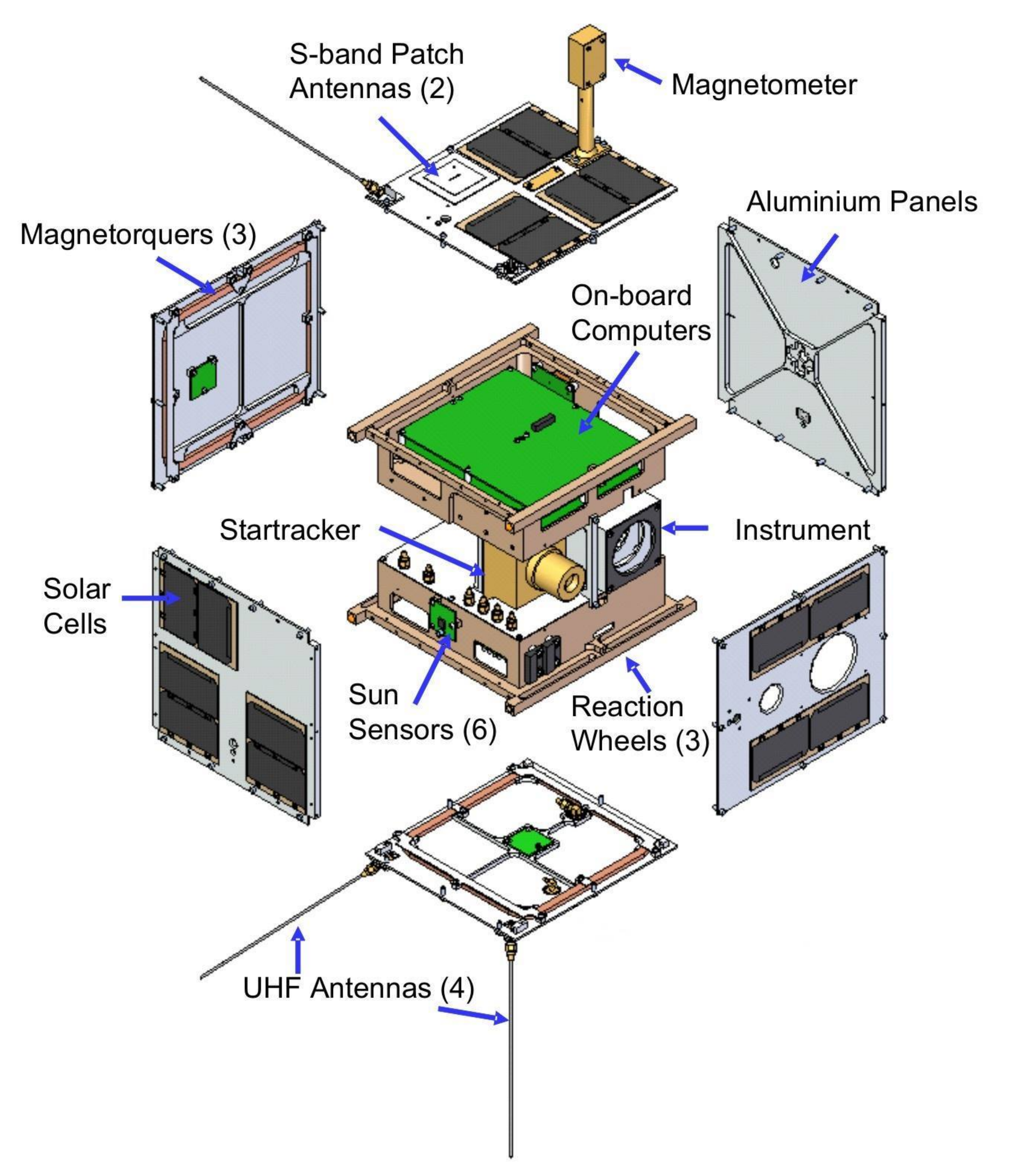}
\caption{\small{A schematic view of the structure of the BRITE spacecraft with the key components labeled.}}
\label{fig:canx3}
\end{center}
\end{figure}

The key components of the BRITE nanosatellite are shown in Fig.~\ref{fig:canx3}. While all six BRITE nanosatellites share the same basic design, on-orbit experience with the first two to be launched (the Austrian nanosatellites) prompted some changes in the designs of the four to follow. 
%

\subsection{Optics}
\label{sect:optics}

Ceravolo Optical Systems\footnote{Ceravolo Optical Systems are located in Ottawa, Canada (see \url{http://www.ceravolo.com}).} provided the two optical designs for the red and blue filters. Both share the same basic configuration, with slight optimizations to ensure that the point-spread function (PSF) of a star image meets the mission requirements. 


The optical cell contains the lenses, spacers, and an O-ring. The inner surface of the cell is a fine-pitch thread which is anodised matte black to dampen internal reflections. The rubber O-ring, located between the last lens and the header tray, where the CCD and focusing mechanism are housed, is preloaded with 200\,N of force. It provides a light-tight seal around the optical components and minimises the risk of damage from vibrations during the launch. The O-ring also absorbs lateral mechanical stress, caused by thermal flexure of the optical components. Along the optical axis thermal flexure is allowed for by  a 0.1\,mm gap between the lens rims and the inner cell wall. The lens configurations are telecentric, meaning that incoming light rays from point sources are collimated with respect to the optical axis. This results in constant magnification for all locations.

The finalised design for the Ceravolo optical setup can be seen in Fig.~9 of Paper I. The lens thicknesses and spacings are slightly different for each bandpass to take into account the difference in indices of refraction. There are five lenses in each configuration made from Schott glass. 
The BRITE optical design is a variation on the standard Double Gauss lens, a well established design for its ability to suppress 4th-order wave aberrations. The net focal length is 70\,mm for the red and blue configurations. 

\begin{figure}[!htbp]
\begin{center}
\includegraphics[trim=8.5cm 7.8cm 6cm 6.3cm,clip,width=0.75\textwidth]{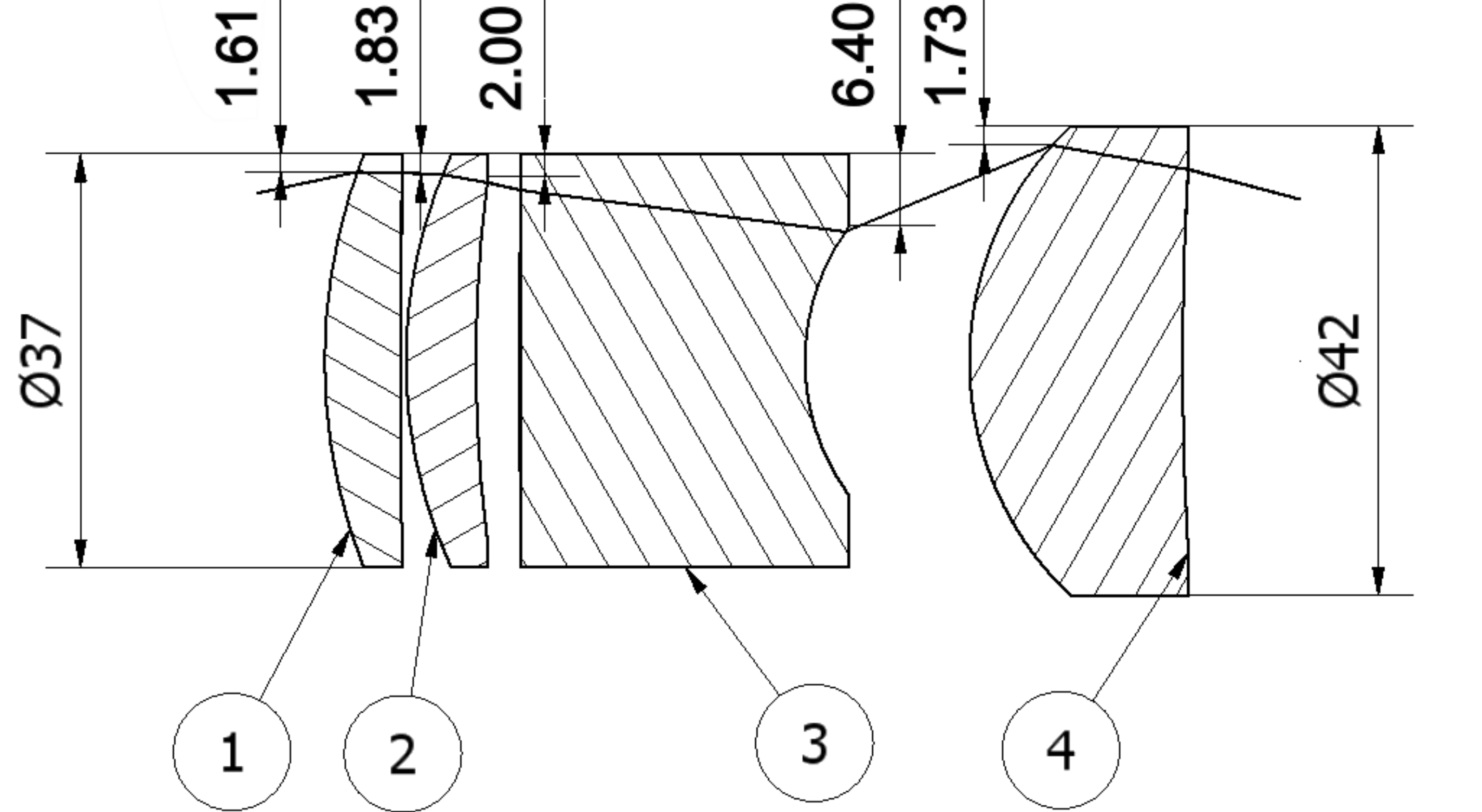}
\caption{\small{BRITE Heweliusz red (BHr) optical design. 
}}
\label{fig:optics_hev}
\end{center}
\end{figure}

The BHr satellite (the last of the BRITEs launched) has a slightly different optical design, with only four lenses. This 4-lens design, developed at the Space Research Center (SRC) in Poland, is shown in Fig.~\ref{fig:optics_hev}.  It produces a more consistent PSF profile across the focal plane than the original Ceravolo design. The field of view, however, is reduced to 24$^{\circ}$ degrees.


\subsection{The BRITE CCD and electronics}
\label{sect:ccd}
\begin{figure}[!ht]
\begin{center}
\includegraphics[trim=0cm 0cm 0cm 0cm,clip,width=0.47\textwidth]{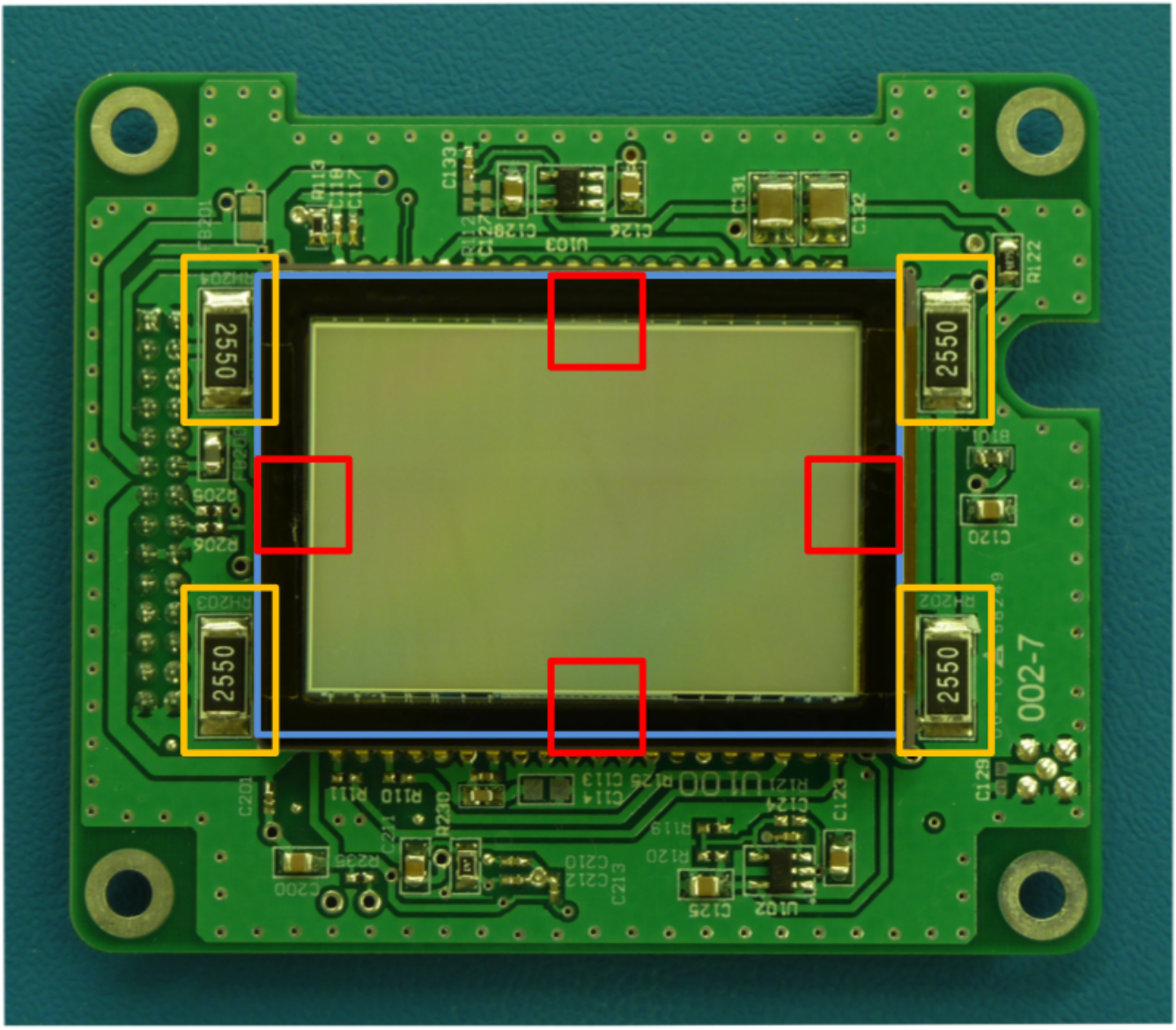}
\caption{\small{A photograph of the CCD header board (courtesy of SFL). The temperature sensors and heating elements are indicated by the red and yellow boxes, respectively.}}
\label{fig:ccd_headerboard1}
\end{center}
\end{figure}

The CCD header tray contains the CCD and thermal-control electronics. These components are mounted on the CCD header board, shown in Fig.~\ref{fig:ccd_headerboard1}. The CCD is located on a separate circuit board external to the telescope. This separation ensures that only a fraction of the heat generated by payload operations impacts the CCD. Furthermore, the CCD is electrically isolated from all other electronics, including the IOBC and the CCD driver preventing electrical interference.

The BRITE CCD is the KAI-11002 CCD from Kodak Truesense Imaging\footnote{Kodak Truesense Imaging is now called ON Semiconductor, after merging and rebranding. Documents relating to the KAI-11002 can be found at http://www.onsemi.com}. It is a front-illuminated, interline-transfer (ILT) device , not requiring a
mechanical shutter to avoid image smearing when read out. Since the rolling electronic shutter function of this CCD is also employed, we do not use the ability of ILT CCDs to take the next exposure while the previous one is read out. The KAI-11002 was chosen over other candidate CCDs because of its affordability, low power consumption, and its exemplary noise and dark current performance at operating temperatures within the anticipated BRITE on-orbit range (20\,$\pm$\,15$^{\circ}$C). The CCD has anti-blooming protection to reduce the
spread of saturated signal to other pixels. 
There are 4072\,$\times$\,2720 total pixels of which 4008\,$\times$\,2672 are active (light collecting) while the outer rows and columns are ``dark'' and ``buffer'' regions. Pixels are arranged in a rectangular matrix pitched 9 $\mu$m apart which combined with the optics gives a plate scale of 27$\arcsec$ per pixel. Full specifications of the CCD are listed in Table 1.

\begin{figure}[!t]
\begin{center}
\includegraphics[trim=1cm 11cm 1cm 11cm,clip,width=0.75\textwidth]{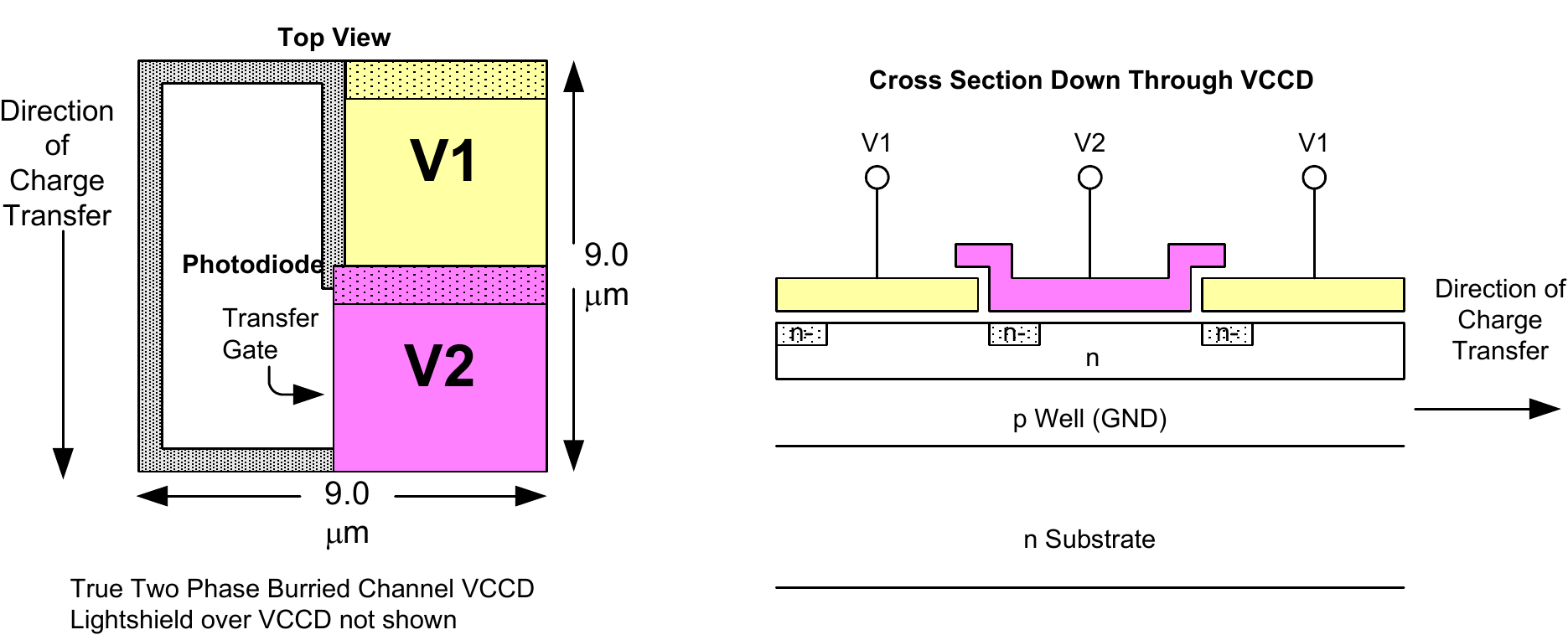}
\caption{\small{The pixel architecture of the two-phase buried channel CCD (courtesy of Kodak Image Sensor Solutions). On the left is a front-facing view, which shows the division between the illuminated portion of the pixel and the vertical transfer buffer. On the right is a horizontal cross-section through the CCD.}}
\label{fig:ccd_pixel1}
\end{center}
\end{figure}

The CCD architecture which enables interline exposures is illustrated in Fig.~\ref{fig:ccd_pixel1}. Each pixel on the CCD is comprised of two halves. The left half (see Fig.~\ref{fig:ccd_pixel1}) contains the silicon-doped photodiode which is fully exposed to incoming photons, while the right half is a vertical transfer region completely shielded by a layer of aluminum. A microlens above each pixel directs most of the incoming photons onto the light-sensitive half, increasing the quantum efficiency from about 16\% to nearly 50\%. 
%
\begin{table}[!ht]
\begin{center}
\caption{\small{Specifications of the KAI-11002 CCD provided by the manufacturer, for a temperature of $40^{\circ}\rm{C}$, which is close to the upper limit on the operating temperature of the BRITE satellites.}}\label{tab:kodak_ccd_specs}
\vspace{3mm}
\centering 
\begin{tabular}{ll}
\hline\hline\noalign{\smallskip}
Parameter & Specification\\
\noalign{\smallskip}\hline\noalign{\smallskip}
Total number of pixels & 4072 (H)\,$\times$\,2720 (V) \\
Active pixels & 4008 (H)\,$\times$\,2672 (V)\\
Chip dimensions & 37.25\,mm\,$\times$\,25.7\,mm \\
Pixel dimensions & 9\,$\mu$m\,$\times$\,9\,$\mu$m \\ 
Imager diagonal & 43.33\,mm \\
Chosen saturation level per pixel & 90,000\,$e^{\minus}$ \\
Quantum Efficiency per pixel & 50\% \\
Total read-out noise per pixel (at 40$^{\circ}$C) & 30\,e$^{\minus}$\\
Worst case cumulative CTE & 93.5\%\\
ADC bit resolution & 14 \\
\noalign{\smallskip}\hline 
\end{tabular}
\end{center}
\end{table}

%


Photo-electrons generated in the silicon-doped photodiodes are held in a potential well, which has a charge capacity dictated by the voltage on the substrate. The maximum charge capacity in the photodiodes corresponds to a low voltage, 8\,V. When the voltage is increased to 40\,V, the photodiodes lose all of their charge capacity and release the electrons into the adjacent shielded vertical register. It is by applying brief pulses of 40 V that exposures are ended; this is the virtual shutter. The charges are rapidly transferred onto the vertical register, thereby avoiding image smear. Once the charge packets are in the vertical register they are shifted (clocked) downwards by alternating the voltages on the electrodes (V1 and V2 in Fig.~\ref{fig:ccd_pixel1}). When they reach the bottom they are shifted along the horizontal register, where the packets are counted and digitised by the analog-to-digital converter (ADC).

The full well potential (saturational level) of the CCD is about 90,000 electrons per pixel. Applying a voltage of 8 V across the substrate does maintain the highest possible dynamic range, but unfortunately provides the minimum anti-blooming protection.
Blooming, caused by excess electrons overflowing the pixel boundaries, can be avoided by releasing excess charge into the substrate before it is transferred into the vertical register. However, this is only possible when the charge capacity of the photodiode is kept lower than that of the vertical register. For the BRITE mission, the saturation level has been set at 60,000 electrons to ensure anti-blooming protection for bright spots 100 times the saturation level.


Dark current and read noise influence the signal-to-noise ratio (SNR) of an
image.  These are temperature dependent, with the dark current doubling
every 6-7$^{\circ}$C increase in temperature, while the total read noise is 30 electrons at 40$^{\circ}$C for an otherwise perfect KAI-11002 device (see Table~\ref{tab:kodak_ccd_specs}).


Lastly, charge transfer efficiency (CTE) is a measure of the number of
electrons lost due to traps caused by impurities in the silicon lattice.
The manufacturer's stated CTE per pixel of the KAI-11002 device is greater
than 0.99999. However, as described below, the CTE of a healthy CCD can be
degraded by radiation damage (see Sect.~\ref{sect:radiation}), which is problematic for imaging from space.
The results of pre-flight tests and predictions  of CCD performance are
presented in Section \ref{sect:ccd_testing}.


%
%

\subsection{Thermal control}
\label{sect:thermal}
The level and stability of the CCD's thermal environment is a major reason explaining why the precursor space photometry missions ensured that their CCD temperatures were maintained at fixed low values (-40$^{\circ}$C  for MOST and CoRoT; -80$^{\circ}$C  for Kepler). Limits of the cost and size of a BRITE nanosatellite meant that even a passive radiator was not an option. 
Instead the CCD is made as cool as possible by mitigating various heat sources, and thermal stability is achieved by the following strategies. First, the CCD header board is isolated from the IOBC and other electronics, mounted on a separate external header board. This reduces the heat dissipation near the CCD. Second, the CCD is in the center of the spacecraft, where thermal fluctuations are smallest.  Third, the CCD temperature is monitored by four sensors and can be adjusted by four trim heaters (see Fig.~\ref{fig:ccd_headerboard1}). Temperature is kept relatively stable by performing ``Sun pointing'' maneuvers when not collecting data, reducing heat absorbed from the Sun (see Sect.~\ref{temp-stability}). 

While the average daily temperature changes gradually during a one or two
month target window, the temperature of each BRITE satellite is modulated
during its orbit by typically 4\,--\,5$^{\circ}$C (see Sect.~\ref{sect:operations}).  The light curves show
these same trends due to changes in pixel sensitivity with temperature. 
While this is unavoidable within the BRITE design,  decorrelation
techniques are employed to reduce this source of systematic error in the affected
light curves. These efforts have been discussed briefly by \cite{pigu16} and will be treated in greater detail in subsequent papers.


\section{Pre-flight testing}
\label{sect:pre_flight_testing}

\subsection{Optical}
\label{sect:opt_simulations}
\begin{figure}[htbp]
\begin{center}
\includegraphics[trim=0cm 0cm 0cm 17cm,clip,width=0.75\textwidth]{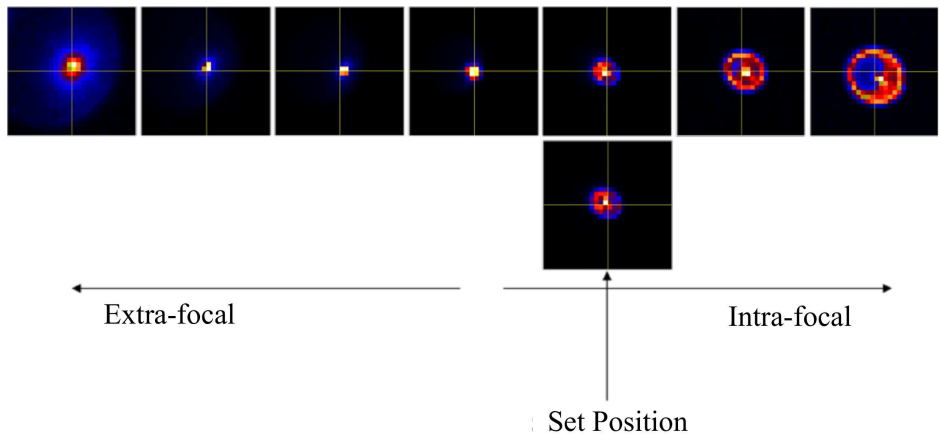}
\caption{\small{The PSF at different offsets from the focal plane, on the optical axis. Each 30 $\times$ 30-pixel subraster shown here was taken after adjusting the CCD position 0.125\,mm further toward (intra-focal) or away from (extra-focal) the optics for a given position on the detector. The intra-pixel set position of 0.0325 mm is shown just below. These results are for UBr, although the outcomes were similar for BTr.}}
\label{fig:red_focus}
\end{center}
\end{figure}

\begin{figure}[htbp]
\begin{center}
\includegraphics[trim=3cm 0cm 2cm 1cm,clip,width=0.75\textwidth]{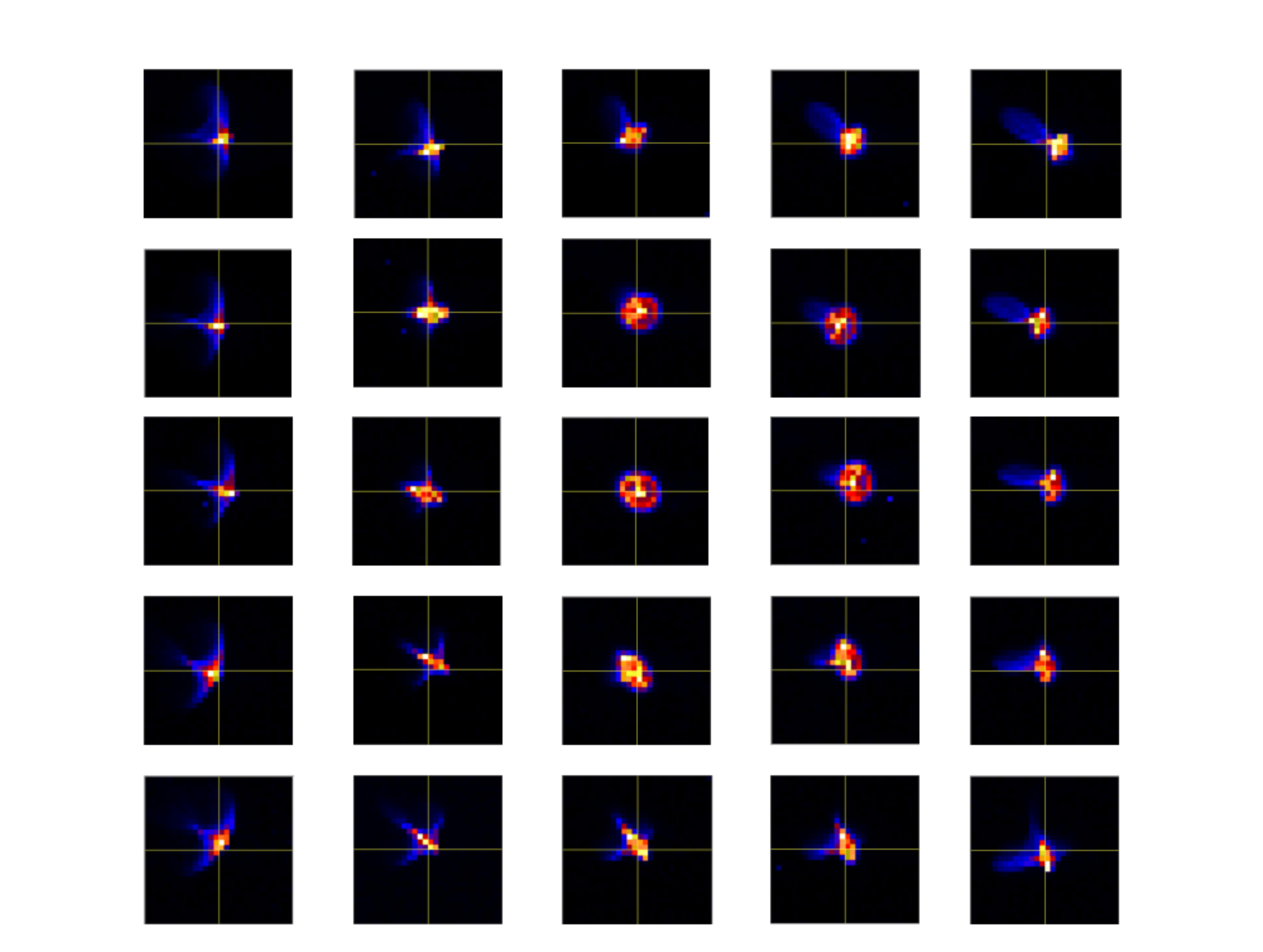}
\caption{\small{Artificial star images obtained with the UBr camera and CCD. To simulate PSF variation as a function of the boresight axis, the telescop was offset in pitch (vertical offset) and yaw (horizontal offset). The central subraster (30$\times$30 pixels) is aligned with the boresight axis and each consecutive subraster represents a shift of 5$^{\circ}$.}}
\label{fig:psftest}
\end{center}
\end{figure}

Pre-flight testing was conducted at SFL, with support from the University of Vienna, mainly to align the CCD with respect to the optics. Test results gave the position of the CCD relative to the focal plane that best produces a smoothly varying PSF across the FOV with the fewest and smallest spikes. Zemax software analyses of a prototype camera produced an initial PSF model for comparison with actual PSF profiles at various separations from the focal plane.
Figure~\ref{fig:red_focus} shows the images used to determine the position of the CCD at which the PSF most closely resembles the model. Spot tests were conducted by shifting the image plane in increments of 0.125\,mm, both toward (intra-focal), and away from (extra-focal) the optics. These tests favored an intra-focal placement of the CCD. Fine-tuning at smaller increments of 0.0325\,mm  demonstrated that a CCD offset of 0.0325 mm produced PSFs with minimum spikes, while ensuring that the light was distributed broadly enough (across a diameter of about 8 pixels) to avoid undersampling. 
Once a good match was found for a PSF in the center of the CCD, the quality of the PSF across the entire FOV was checked, as shown in Figure~\ref{fig:psftest}.  Even the best solution is not ideal, as it produces spiky, irregular shaped PSFs on the edges of the CCD  making uniform flux measurements more difficult.

On-orbit images with the BRITE optics show PSFs which are more uniform across the FOV than expected from the pre-flight tests.  Of the five functioning BRITEs, UniBRITE red (UBr) exhibits the largest differences in PSF profile across the FOV. While the exact reason for this is unknown, it is likely due to small errors in the centering and positioning of the lenses.

%

\subsection{CCD}
\label{sect:ccd_testing}
The predicted performance of the CCD of each BRITE was assessed through a
series of pre-flight acceptance tests in accordance with the BRITE
Instrument Acceptance Test Plan.  This document details procedures for appraising bias level, dark current, read noise,
gain, linearity and dynamic range, and the results are presented in the
Payload Instrument (Flight) Acceptance Test Results document for each
BRITE.  The procedures are further described by \cite{cheng12}. Tests were performed on all six flight CCDs at four operating temperatures: 0$^{\circ}$, 10$^{\circ}$, 20$^{\circ}$ and 30$^{\circ}$C.  Survivability tests at -20$^{\circ}$
and 60$^{\circ}$ C were executed to test the limits of the device.  Each test
involved taking a series of images in one or more of the following
categories: bias, dark, and gradient (described below). Images were analyzed with IRAF\footnote{IRAF is distributed by the National Optical Astronomy
Observatories, which are operated by the Association of Universities for
Research in Astronomy, Inc., under cooperative agreement with the National
Science Foundation}  and DS9 software.  Summaries of the tests
and their results (which were comparable for all six devices) are presented below.


\subsubsection{Bias}
\label{sect:bias}

The bias level of the CCD is a preset value added to the electronic signal of each pixel
during readout, typically between 70 and 130\,ADU (analog-to-digital units; conversion to electrons given in Sect.~\ref{sect:gain}). This
is just sufficient to overcome systemic noise and ensure there are no
negative values at readout that could be misinterpreted by the ADC. The mean
bias offset can be adjusted in flight.  The
test described here was used to assess the stability of the bias level,
throughout continuous operation, and over the full range of operating
temperatures.  It was also used to confirm that the inherent electronic
noise follows a Gaussian distribution.

\begin{table*}[!ht]

\caption{\small{Bias and stability requirements and results from pre-flight
tests for all BRITE satellites.}\label{tab:bias}}
\vspace{3mm}
\begin{tabular}{p{0.5\textwidth}p{0.44\textwidth}}
\hline\noalign{\smallskip}
\multicolumn{1}{c}{Requirement} & \multicolumn{1}{c}{Result} \\
\noalign{\smallskip}\hline\noalign{\smallskip}
Mean bias level = 100\,$\pm$\,30\,ADU never falling below 40\,ADU  & The mean
bias level is 100\,$\pm$\,30\,ADU at $T\le\mbox{30}^{\circ}$C, but exceeds 500\,ADU at
60$^{\circ}$C \\ 
\noalign{\smallskip}\hline\noalign{\smallskip}
The bias of any given pixel should not vary from the mean level by more than 2 ADU. & 95\% of pixels
are within $\pm$2\,ADU of the mean. 100\% of
pixels are within $\pm$3\,ADU of the mean.\\
\noalign{\smallskip}\hline\noalign{\smallskip}
Distribution of values should be Gaussian. & The
distribution is Gaussian.\\
\noalign{\smallskip}\hline
\end{tabular}
\end{table*}

A bias frame is an image taken in the dark with a zero-second exposure time. 
This frame captures the characteristics of the electronic noise, as there
is no light source to contribute shot noise and no time to accumulate
thermally generated electrons (dark current).  Electronic noise exists
because of manufacturing irregularities and because the ADC is imperfect.  By analysing the bias frames taken at various
temperatures over an appropriate length of time, the electronic noise can be
characterised and the stability of the system can be measured.

A series of bias frames was taken at the temperatures previously stated,
over one hour at each temperature. Subsections of each frame were analysed. The results are compared to the requirements in
Table~\ref{tab:bias}.

\subsubsection{Gain}
\label{sect:gain}
\begin{figure}[!htbp]
\begin{center}
\includegraphics[trim=0.5cm 0cm 0cm 0cm, clip, width=1.0\textwidth]
{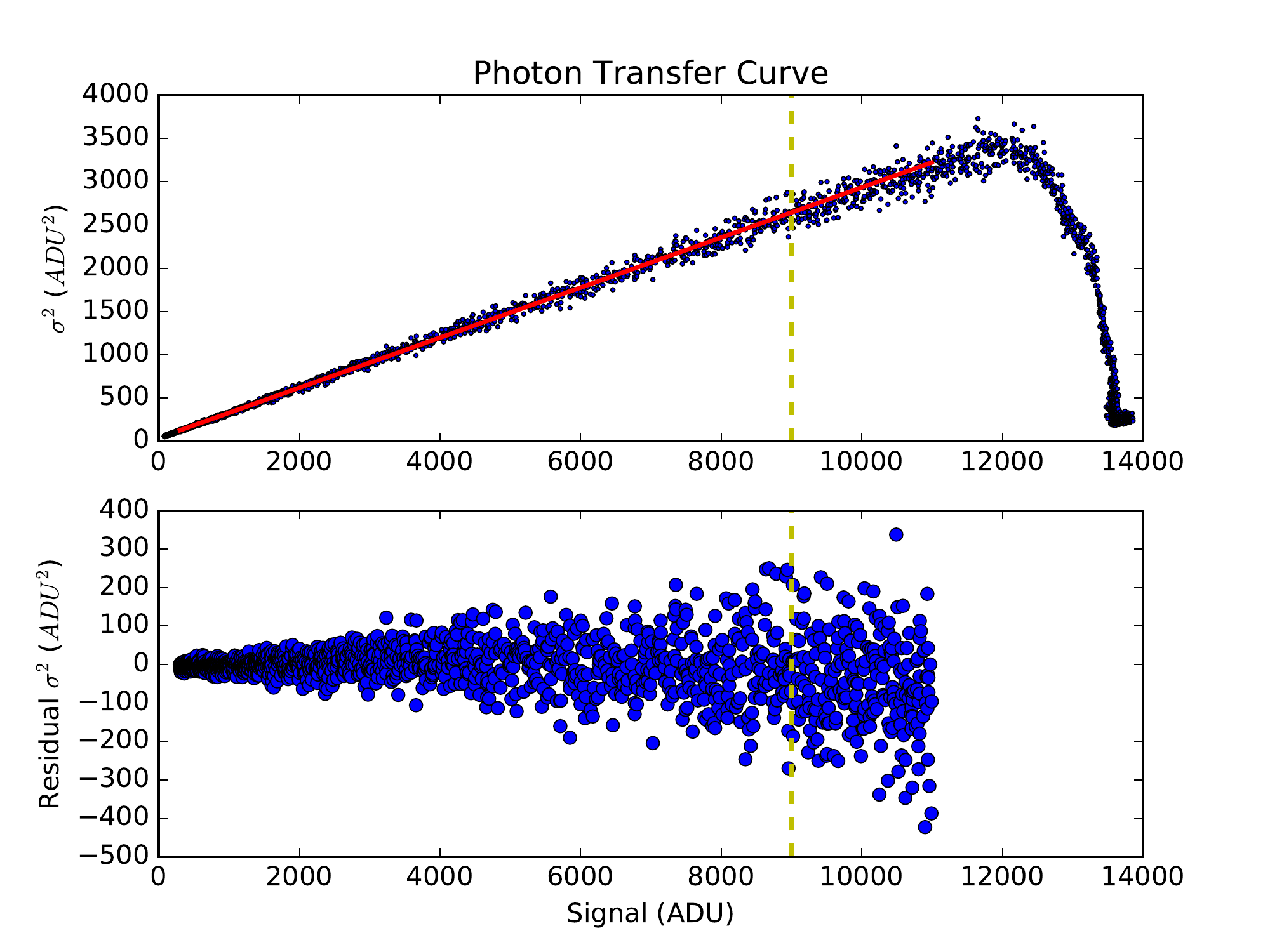}\caption{\small{CCD linearity curve at $20^{\circ}$C (top) with the dashed yellow line  marks the high end of the linear regime. the red line showing the fit to the linear portion and extended to 11000 to highlight how the residuals (bottom) change. The y intercept is the square (in $\rm{ADU}^{2}$) of the readout noise. }}
\label{fig:ccd_linearityCurve}
\end{center}
\end{figure}

Gain is the conversion factor from photo-electrons to ADU, which takes place during readout in the ADC.  The gain can be directly measured using a precision photon source, but this was beyond BRITE's budget.  Instead, the gain was estimated by measuring the variance of the system at different light levels
and plotting the results versus mean signal level (in ADU). 
Least squares fitting of the data gives the slope of the photon-transfer curve (see Fig.~\ref{fig:ccd_linearityCurve}). 
 
Gradient images were obtained by exposing the CCD to a single LED light
source through a neutral density filter and a single lens.  The LED was
placed at one side of the CCD, so that the furthest pixels would be
sufficiently dark and the illumination roughly constant along columns. 
These images contain the necessary information regarding the dynamic range
(from bias level to saturation), provided the CCD is illuminated with a
gradual transition of photon intensity.  Mean dark images (see Sect.~\ref{sect:dark}) at the same
temperature and with the same exposure time (0.75 s) were subtracted from
each gradient image.  The net gradient images were scaled to the same mean
value, and then mean images and sigma images, created by calculating the standard deviation associated with each pixel based on the list of raster inputs,  were formed from this set.  A boxcar filter
with dimensions 1x100 pixels smoothed the mean and sigma images along the
columns. These data were then plotted as mean signal versus variance, where the variance is calculated by squaring the sigma values ($\sigma^2$). 
These photon-transfer plots, like the one in
Fig.~\ref{fig:ccd_linearityCurve}, were constructed for each temperature
tested.  They were used to determine the system gain using the relation,
\begin{equation}
\label{eq:gain}
\sigma^{2} = \frac{1}{K} S + R^{2},
\end{equation}
where $\sigma$ is the standard deviation ($e^{-}$)  at signal $S$ (ADU), $K$ is the gain (ADU/$e^{-}$), and R is the readout noise ($e^{-}$).  In practice, we multiply equation (1) through by $K^{2}$ and
plot the results as variance in $\rm{ADU}^{2}$ versus signal in ADU.  The gain $K$ is simply the slope of the linear portion of the plot of the total
variance versus the signal.  It is not affected by small pixel to pixel gain
differences (flatfield noise) when the variance is measured by differencing two exposures of identical exposure times.  The results of this test are shown in Table~\ref{tab:gain} .  Variations in the values at different temperatures are most likely a result of error propagation. A value of 3.5 $e^{-}$/ADU was adopted as the inverse gain before launch for all BRITEs.
\begin{table}[h!]
\centering
\caption{\small{Inverse gain values. All values have a standard deviation of about 0.1
among all 6 flight CCDs tested at each temperature}\label{tab:gain}}
\vspace{3mm}
\begin{tabular}{rrrrrrr}
\hline\noalign{\smallskip}
Temperature ($^{\circ}\rm{C}$) & \minus 20 & 0 & 10 & 20 & 30 & 60 \\
Inverse Gain ($\rm{e^{-}ADU^{\minus1}}$) & 3.43 & 3.42 & 3.49 & 3.44 & 3.40 & 3.60 \\
\noalign{\smallskip}\hline
\end{tabular}
\end{table}

\subsubsection{Saturation and linearity}
Saturation and linearity tests set the upper limit in photon intensity which can be measured reliably at a given exposure time; i.e. at what level the signal will deviate unacceptably from linearity. This
information is then used to select target stars and exposure times based on stellar apparent magnitude.

In theory, having a full-well capacity of 60,000\,$e^{\minus}$ should mean
that the system is linear for all measurements below this limit, i.e.,  one
photon absorbed in the silicon produces one photo-electron.  In practice the system will deviate from linearity before this limit is reached, because near the
saturation level the pixels have a reduced probability of capturing
photons.  The point at which the deviation exceeds an acceptable
level is the ``linear full-well capacity''.  It is essential to carefully
establish both the full-well capacity and the point of deviation from
linearity, to enable correct interpretation of the resulting data.

Saturation occurs at the maximum signal level that contributes to a positive slope in the photon-transfer
curve (see Fig.~\ref{fig:ccd_linearityCurve}).  For temperatures up to 60$^{\circ}$ C the linear full well capacity is above the 30,000\,$e^{-}$ (8,600\,ADU) required by the BRITE mission.

The point at which the system deviates from linearity can also be extracted
from the photon-transfer curves, by analysing where the data depart from
the trend.  The deviation from linearity actually starts at around
9,000\,ADU, as can be seen in Fig.~\ref{fig:ccd_linearityCurve}.  
This means any value $\le$\,9,000\,ADU is a linear measurement and 
can be directly related to photo-electrons using the gain value.  
Any values significantly greater than 9,000\,ADU are less
reliable and should be treated with caution, especially at higher operating
temperatures. It is important to note that while this value was well known before launch, the CCD clock frequency has been changed during the post launch phase of the mission (see Sect.~\ref{sect:chop}) so verification must be done on data collected beyond this point to ensure the relationship has not changed.  


%
\begin{table}[h!]
\centering
\caption{\small{Dark current statistics for the full range of operating
temperatures and exposure times tested. `sat' indicates where the pixels with the highest sensitivity have already reached saturation, making accurate calculation impossible.}}\label{tab:darkcurrent2}
\vspace{3mm}
\begin{tabular}{crrrrrr}
\hline\noalign{\smallskip}
{Temp.~($^{\circ}$C)} & \multicolumn{6}{c}{Dark current
($\rm{e}^{-}\rm{s}^{\minus1}\rm{pixel}^{\minus1}$)}\\
\noalign{\smallskip}\hline\noalign{\smallskip}
  & $1\,\rm{s}$ & $3\,\rm{s}$ & $10\,\rm{s}$ & $30\,\rm{s}$ & $60\,\rm{s}$ & $90\,\rm{s}$\\
\noalign{\smallskip}\hline\noalign{\smallskip}
$\minus$20 & 0 & 0 & 0 & 0 & 0 & 0\\
0 & 2 & 2 & 2 & 3 & 3 & 4\\
10 & 8 & 8 & 8 & 8 & 8 & 8\\
20 & 21 & 21 & 21 & 21 & 21 & 21\\
25 & 35 & 35 & 34 & 34 & 33 & 33\\
30 & 49 & 49 & 49 & 50 & 50 & sat\\
60 & 534 & 534 & sat & sat& sat& sat\\
\noalign{\smallskip}\hline
\end{tabular}
\end{table}

\subsubsection{Dark current}
\label{sect:dark}
Dark current is the rate (in e$^{\minus}$\,s$^{\minus1}$\,pixel$^{\minus1}$) at which
thermally-excited electrons are produced in the CCD at a given temperature. Even
though the KAI-11002 was chosen for its  low dark current, it was
still important that the dark current was characterized pre-flight at the full range of
operating temperatures anticipated for the BRITE satellites.

A series of three dark images was taken at each of the 
temperatures and exposure times in Tab.~\ref{tab:darkcurrent2}. Several bias frames (0 second exposures) were taken at each temperature, and their means were subtracted from the dark
images. The mean dark images were analysed, first to check that the dark current rates and distributions were within mission requirements, and second to determine the relationship between dark current rate and exposure time. 
The Kodak CCDs are well known to exhibit a non-uniform dark current distribution, with a few ``hot'' pixels carrying much of the overall dark current.  Analysis of histograms of the dark current showed that pre-flight typically for a 10 second exposure at room temperature (22-23$^{\circ}$C) the median value was near 0, with 5\% of pixels above 7 ADU, 1\% above 13 ADU, 0.5\%
above 18 ADU, 0.1\% above 200 ADU and 0.05\% above 400 ADU (multiply by 3.5 to obtain electrons). The hottest pixels can be avoided or excluded in data
analysis. The degradation of this excellent pre-light performance after launch is discussed in Section \ref{sect:radiation}.

Table \ref{tab:darkcurrent2} shows the \textit{average} dark current as a function of
temperature and exposure time. The average dark \textit{current} does not change
with exposure time, but displays the customary exponential increase with
increase of temperature. Note that most of the dark current is concentrated
into a small fraction of pixels in the operating range of 0-30$^{\circ}$C.

\subsubsection{Readout noise}

Readout noise is added to an image when each charge packet (pixel) is read out  and converted to an amplified voltage which is then digitized. This noise  is dependent on CCD clocking time, the time required to read out each pixel. A faster readout comes at the cost of 
proportionally greater noise (amplifier Johnson noise).  It is important
that the readout noise be Gaussian (i.e.,~purely random), so that it can be
reduced  by  $\sqrt{n}$ for every $n$ images combined.  
Readout noise is independent of exposure time and mostly independent 
of operating temperature, but at high temperatures thermal noise may interfere with the measurements.

The total readout noise for the system was measured in units of e$^{\minus}$\,pixel$^{\minus1}$ and tested for a Gaussian distribution.  Pairs of bias frames were subtracted from each other and the pixel values used to create a histogram, with the assumption that the value of one pixel over several images has the same distribution as many pixels in one image. After confirming this histogram was Gaussian, the standard deviation of the histogram was measured and scaled down by $\sqrt{2}$ to account for the variance being effectively doubled by co-subtraction.  The readout noise for the whole system is equal to the scaled standard deviation. The requirement for the readout noise is
$<$\,30\,e$^{\minus}$\,pixel$^{\minus1}$ for all operating temperatures
($\le$\,30$^{\circ}$C).  The corresponding results are
presented in Table~\ref{tab:readout}.
\begin{table}[h!]
\centering
\caption{\small{Readout noise. These values have a standard deviation of about 2 electrons per pixel over all six CCDs at $25^{\circ}$ C.}} \label{tab:readout}
\vspace{3mm}
\begin{tabular}{rrrrrrrr}
\hline\noalign{\smallskip}
Temperature ($^{\circ}$C) & $\minus$20 & 0 & 10 & 20 & 25 & 30 & 60 \\
Readout noise (e$^{\minus}$\,pixel$^{\minus1}$) & 13 & 13 & 14 & 16 & 18 & 19 & 63\\
\noalign{\smallskip}\hline
\end{tabular}
\end{table}

\subsubsection{Solar Illumination Safety}
\label{sect:sunstare}

Lacking a shutter, BRITE cannot protect the CCD from accidental exposure
to the Sun through the telescope.  Preliminary calculations suggested that
such accidental exposure would not harm the detector, but a practical
sun-stare test was conducted using a non-functioning mechanical model of the
KAI-11002 CCD \citep{dwyer2007}.  A lens with similar aperture and
focal length to those of BRITE was used to focus the Sun with the aid of a
heliostat.  Temperature sensors were used to measure the temperature of the
model CCD in various places. 

The model CCD and thermal sensors were placed inside a bell jar evacuated to
150 mTorr, and a number of tests were conducted.  The substrate of the chip
was found to be sufficiently conductive to maintain temperature uniformity
to within a fraction of a degree when the chip was illuminated with a focused
image of the Sun, and the increase of temperature under constant
illumination never exceeded 37$^{\circ}$C (6$^{\circ}$C above the heater-maintained ambient)
in one hour of sun-staring, not including heat loss due to conduction to the
rest of the spacecraft.  It was concluded that even four hours of sun
staring would not cause the CCD to exceed the maximum allowable temperature
of 70$^{\circ}$C.  The tests were conducted with wider filter passbands than those of
the actual BRITE B and R filters, or no filter at all.  A pyrometer was used
to measure the solar energy passing into the bell jar .

\section{Launches and commissioning}
\label{sect:comm}

The constellation of BRITE satellites was delivered into orbit via four separate launches, over 18~months. The staggered launch sequence was a byproduct of when funding was received from the participating agencies and the logistics of finding low-cost launch opportunities as secondary payloads, but also minimized the risk of a total loss of the mission at launch. After each satellite was deployed from its launch vehicle, contact was made with one of three ground stations and commissioning begun with procedures developed at SFL. (See below, and Paper I, for additional details). The two Austrian BRITEs were launched first, and so their commissioning lasted the longest (about 8 months). Armed with that experience, the BRITE team was able to commission the subsequent Polish and Canadian satellites more efficiently. The BRITE Toronto red (BTr) nanosatellite was in normal science operations only 8 days after launch. 
\subsection{Launches and orbits}
\label{sect:launch}
The details of the launches and orbits of the BRITE nanosatellites are described below, in the order the satellites were commissioned, with key information summarized in Table~\ref{tab:launches}. All BRITE satellites are in low-Earth orbits (LEO) with periods of about 100 minutes (frequencies near 14 cycles per day).

%
\begin{table*}[!ht]
\centering
\caption{\small{Launch and orbit specifications for all BRITE satellites. Information in the last three columns are provided by the US Space Surveillance Network, operated by US Air Force Space Command (http://www.n2yo.com). BMb never made contact and so no information is known.}}
\vspace{3mm}
\resizebox{\columnwidth}{!}{
\label{tab:launches}
\begin{tabular}{ccccccc}
\hline\noalign{\smallskip}
ID & Full name & Launch date & Operating station & Altitude ($\rm{km}$) & Inclination & Period ($\rm{min}$)\\
\noalign{\smallskip}\hline\noalign{\smallskip}
UBr & UniBRITE & 25 Feb 2013  & Toronto (SFL)  & $775\,-\,790$ & $98.6^\circ$ & 100.4\\
BAb & BRITE-Austria & 25 Feb 2013  & Graz (TUG) & $775\,-\,790$ & $98.6^\circ$ & 100.4\\
BLb & Lem & 21 Nov 2013 & Warsaw (CAMK) & $600\,-\,890$ & $97.7^\circ$ & 99.6\\
BTr & BRITE-Toronto & 19 Jun 2014 & Toronto (SFL) & $620\,-\,643$ & $97.9^\circ$ & 98.2\\
BMb & BRITE-Montr\'eal & 19 Jun 2014 & Toronto (SFL) & --- & --- & --- \\
BHr & Heweliusz & 19 Aug 2014 & Warsaw (CAMK) & $612\,-\,640$ & $98.0^\circ$ & 97.1\\
\noalign{\smallskip}\hline\noalign{\smallskip}
\end{tabular}
}
\end{table*}

\subsubsection{UniBRITE (UBr) and BRITE-Austria (BAb)}

The Austrian satellites, UBr and BAb, were launched on 25 February 2013 aboard a C20 Polar Satellite Launch Vehicle (PSLV), operated by the Indian Space Research Organization (ISRO). Both satellites were placed into Sun-synchronous, dawn-dusk orbits. UBr, built and tested at SFL in Toronto, is operated through the ground station there\footnote{At the time of writing, operation of UBr is being transferred to Graz}. BAb, assembled and tested at the Technical University of Graz (TUG), is operated from the Institute for Communication Networks and Satellite Communication (ICNSC) in Graz, Austria.

The dawn-dusk Sun-synchronous orbit keeps the satellites over the Earth's terminator and yields nearly continuous illumination for the solar panels. However, between November and February, when the Earth's northern hemisphere is most strongly tilted away from the Sun, these highly-inclined orbits will dip briefly (for about 20 minutes) into the Earth's shadow once per 100-min orbit. During this ``eclipse season", the satellites must draw on battery power. The Austrian BRITEs are equipped with AA-MST star-trackers which cannot operate fully when the satellites are too deeply in eclipse. This introduces some short gaps in the Nov-Feb observing sequence of UBr and BAb. 

\subsubsection{BRITE Lem (BLb)}

BLb was launched on 21 November 2013 from Yasny, Russia, aboard a Dnepr rocket operated by the International Space Company (ISC) Kosmotras.  The first of the Polish BRITEs is named in honor of Polish author Stanis{\l}aw Lem.


Unlike the PSLV-C20, which launched the first two BRITEs and stabilizes in orbit before deploying its payloads, the Dnepr rocket starts releasing satellites immediately after reaching the intended altitude while its thrusters are still firing. As a result, BLb was placed in a more eccentric orbit whose perigee is determined by the point of detachment from the rocket. With multiple payloads, as in the case of this launch, the later the satellite is released, the more eccentric is its orbit. The perigee of the BLb orbit is 600 km; its apogee is 890 km. As a result, the ascending node of the orbit precesses by about 100 minutes per year. This means the BLb orbit is not strictly Sun-synchronous and experienced eclipses every orbit when launched. This orbit has evolved over time and is currently Sun-synchronous, which will last until April 2017\footnote{This corresponds to the middle of observation field 17, Carina}. After that, BLb will again undergo eclipses every orbit. 
 
\subsubsection{ BRITE Toronto (BTr) and BRITE Montreal (BMb)}
The Canadian satellites, BMb and BTr, were launched together on 19 June 2014 from Yasny (Russia), onboard a Dnepr rocket operated by the ISC Kosmotras. BTr was quick to establish communication with the ground station at SFL in Toronto, verifying its healthy state.
BTr has a much less eccentric orbit than BLb (despite being launched by the same type of rocket) with perigee 620 km and apogee 643 km. Like BLb, BTr experiences daily eclipses, except when its orbit happens to align with the terminator. 

No contact has been made with BMb and it is believed that the satellite never detached from the third stage of the launch vehicle rendering it unusable.

\subsubsection{BRITE Heweliusz (BHr)} 
Named after Polish astronomer Jan Heweliusz, BHr was the last of the BRITE satellite constellation to be launched.  It was placed into orbit on 19 August 2014 from the Taiyuan Satellite Launch Centre  (China) aboard a Long March 4B rocket operated by the China Aerospace Science and Technology Corporation. BHr has a nearly circular, Sun-synchronous orbit with an altitude of approximately 635\,km. 


\subsection{Commissioning tasks and schedules}
\label{sect:timelines}
\begin{figure}[!h]
\centering
\includegraphics[trim=1cm 12cm 4.5cm 1cm,clip,width=1.0\textwidth]{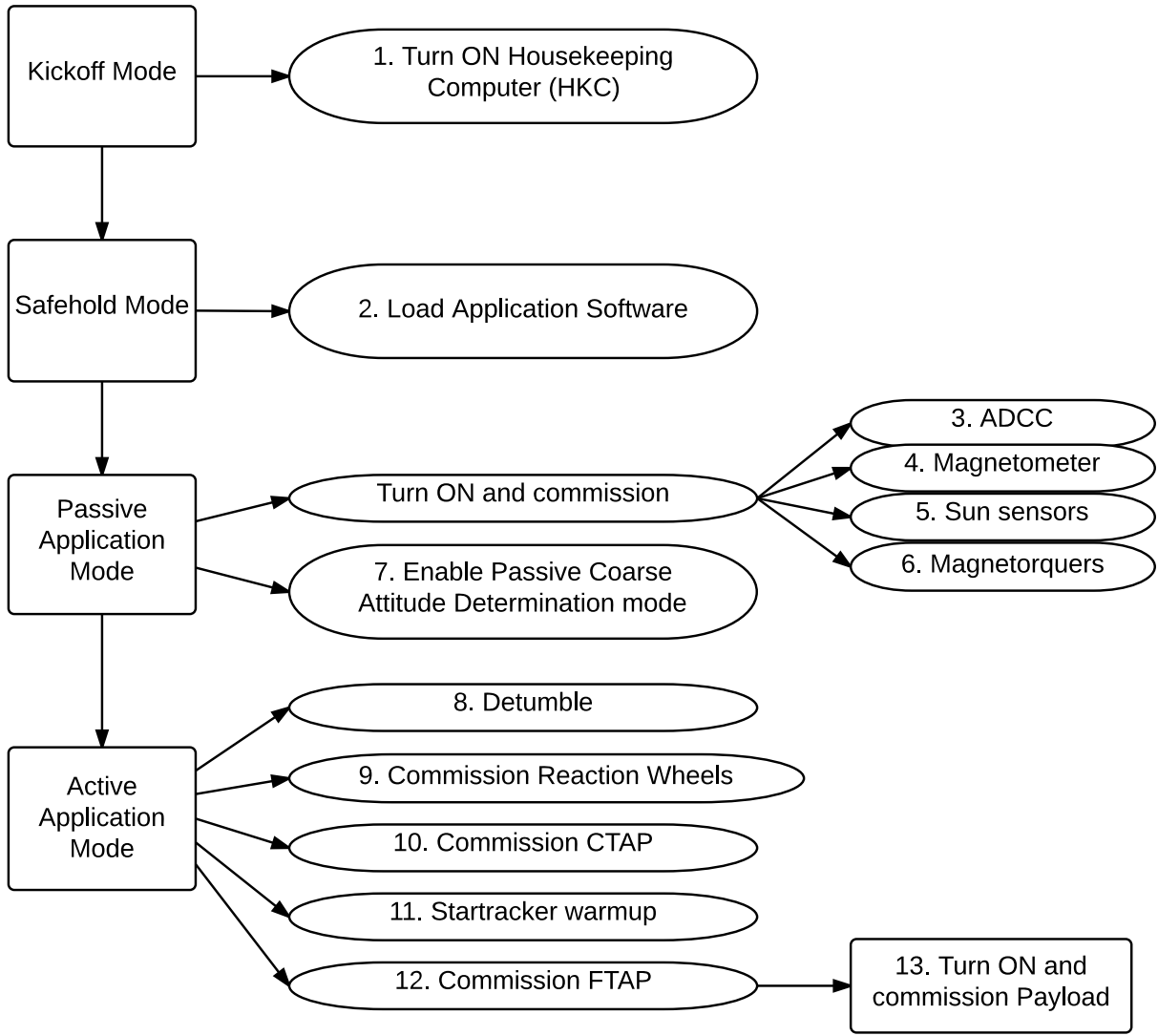}
\caption{\small{The standard commissioning procedure for a BRITE satellite. CTAP and FTAP refer to ``coarse'' and ``fine'' three-axis pointing, respectively.}}
\label{fig:sfl_commission}
\end{figure}

When the satellites were released from their launch vehicles they were automatically in ``kickoff'' mode, during which only communications and power are active. 
Commands relayed by a ground station activate the onboard housekeeping computer (HKC), placing the satellite into ``safe hold'' mode. Then the commissioning begins, with steps shown in Fig.~\ref{fig:sfl_commission}.  All five BRITEs were commissioned in this way, but the exact sequence and time for each step varied from satellite to satellite.
%

\subsubsection{UBr and BAb}
\label{sect:com_ubr}

The first two BRITE satellites placed into orbit (in late February 2013) were the prototypes for the hardware, software and the on-orbit commissioning process.  The commissioning lasted 8 months, ending in October 2013. Stages 1 through 8 (Fig.~\ref{fig:sfl_commission}) were completed in the first month.  Most of the remaining 7 months were occupied with recalibration of the star-trackers without support from its manufacturer. The other main tasks were stages 4 (magnetometer) and 5 (Sun sensors) . 
The five main problems encountered during the commissioning of UBr and BAb were as follows:

\noindent Problem 1: The AeroAstro miniature star tracker (AA-MST). From the outset the AA-MST had various performance issues preventing the satellites from entering, and maintaining fine three-axis pointing (FTAP), including:
\renewcommand{\theenumi}{\alph{enumi}}
\begin{enumerate}
\item{Failure to transfer attitude solutions to the ADCS.}
\item{Failure to calculate reliable quaternions\footnote{A quaternion is a set of four numbers 
(composed of a vector and a rotation about that vector) which uniquely defines the orientation of an object relative to a fixed coordinate system.} outside of a narrow viewing zone, despite the product being rated by the manufacturer as capable of ``all-sky'' coverage. It was found to produce reliable quaternions only when directed at dense star fields in the Galactic plane. This remains an ongoing problem limiting the sky coverage of UBr and BAb.}
\item{Fine pointing drop-outs when passing through the South Atlantic Anomaly (SAA), caused by the increased flux of energetic particles on the star tracker CCD. These confuse the star tracker, by temporarily mimicking stars in the FOV.}
\end{enumerate}

\noindent \textit{Solutions}: Revised software uploaded to the satellites repaired the link between the ADCS and the star tracker. Without manufacturer input and expertise, no clear explanation is available for the poor performance beyond specific parts of the sky. Trial-and-error testing led to an empirical solution, reconfiguring the settings of the star tracker. Both satellites can now observe to within 20$^{\circ}$ of the Moon. The loss of fine pointing in the SAA remains an issue, but this affects only some targets for a maximum of 5 of 14 orbits daily.

\noindent Problem 2: Mis-calibration of the magnetometer. 

\noindent \textit{Solution}: On-orbit recalibration of the magnetometer resulted in better performance, allowing for the actual space environmental conditions and the influences of all running spacecraft subsystems on the local magnetic field. 

\begin{figure}[!ht]
\centering
{\includegraphics[trim=8cm 11cm 7cm 10.6cm,clip,width=0.35\linewidth]{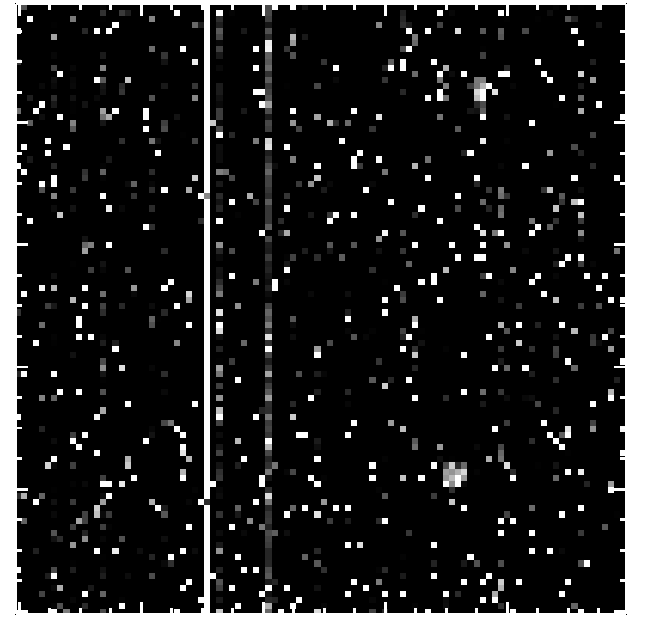}\label{fig:fl_ub}}
{\includegraphics[trim=8cm 11cm 7cm 10.6cm,clip,width=0.35\linewidth]{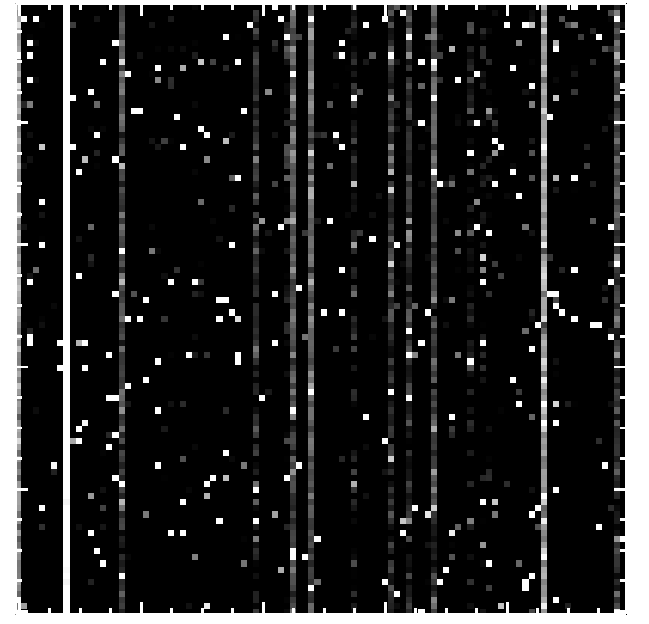}\label{fig:fl_ba}}\\
{\includegraphics[trim=8cm 11cm 7cm 10.6cm,clip,width=0.35\linewidth]{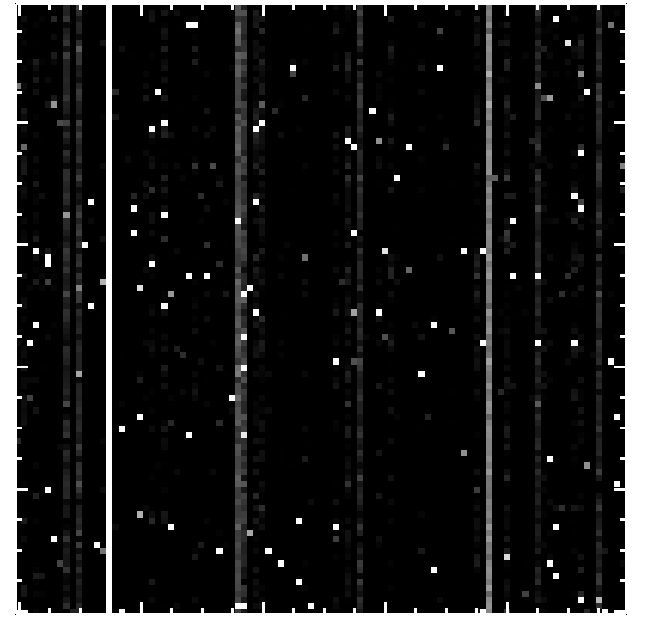}\label{fig:fl_bl}}
{\includegraphics[trim=8cm 11cm 7cm 10.6cm,clip,width=0.35\linewidth]{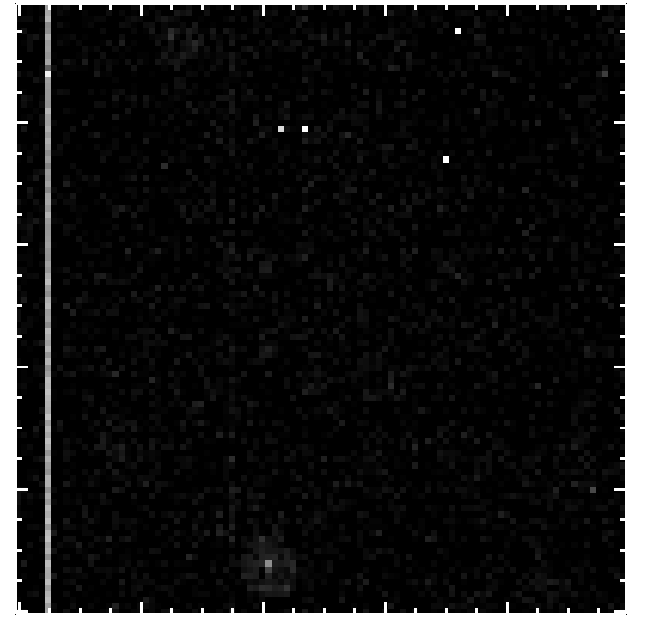}\label{fig:fl_bt}}\\
{\includegraphics[trim=8cm 11cm 7cm 10.6cm,clip,width=0.35\linewidth]{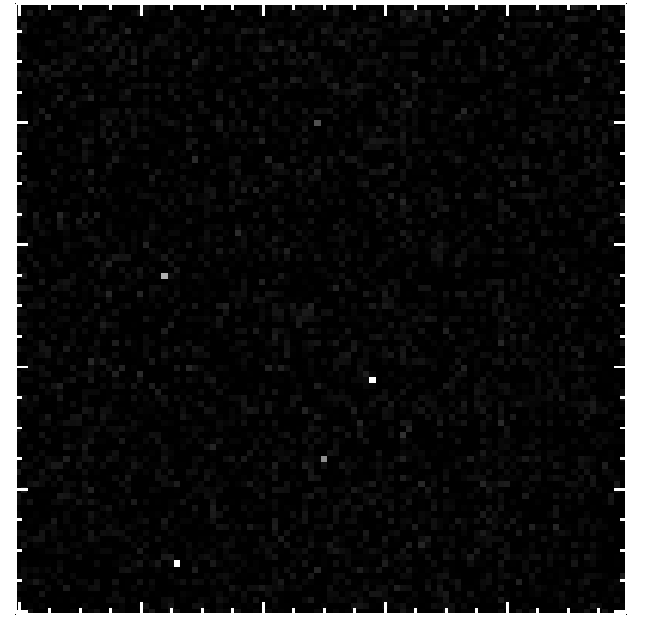}\label{fig:fl_bh}}
\caption{\small{Subrasters of the first full-frame images, obtained during commissioning of each satellite. Top row: UBr (left) and BAb (right), middle row: BLb (left) and BTr (right), and bottom row: BHr. Each subraster covers 100\,$\times$\,100\,pixels from the middle of the CCDs. They show a range of CCD defects caused by radiation damage to the spacecraft during their initial weeks in orbit. Due to differences in the commssioning time scale the first image was taken after several months for BAb, UBr, and BLb, as opposed to only a week for BTr and BHr which also benefited from design modifications before launch, including extra shielding. The significantly reduced radiation damage in BTr and BHr is likely due to both these factors.}}
\label{fig:firstlight}
\end{figure}

\noindent Problem 3: Radio interference with the European BRITE ground stations. Beginning in October 2013, ultra-high frequency (UHF) interference from unknown sources has interfered with BRITE uplinks of commands and software updates and downlinks of data.  This significantly slowed the commissioning. The BRITE mission is not the primary owner of the frequency it uses, so it cannot protest to international frequency regulation agencies. The satellites operated through the Toronto SFL ground station are not affected.  
\paragraph{\sl Solutions:}
\begin{enumerate}
\item{The Toronto SFL station has been used to clear backlogs of data stored on BAb and BLb, so science operations were not interrupted. It was necessary, however, to reduce the number of targets monitored by BAb to the 15 brightest stars in its field. UBr was able to observe all of its 30 targets in the Centaurus field.}
\item{New downlink software tested at SFL was uploaded to all satellites in June 2014. The new protocol does not require the satellite to receive a command to begin downloading to a station. Not only did this improve the situation for the European stations, but it also improved the downlink rate from UBr at SFL.  As of October 2014, SFL downlink rates had increased by almost 70\% (30 to 50 MB/day), well above the original mission requirement.  Because of this, observing windows have been extended to 40 min/day for 30 targets, and an extra field can be observed between primary fields (see Sect.~\ref{sect:operations}).}
\item{Counter-measures have been implemented at the ground stations in Graz (Austria) and Warsaw (Poland) to improve efficiency:
\begin{itemize}
\item Installation of new higher-performance antennas and more precise rotators
\item Nadir tracking during ground contacts to improve link conditions
\item Automatic data rate adjustments in response to changing the link conditions 
\item Packet size optimizations, to reduce the need for re-transmissions
\item Increased manual interactions during operations.
\end{itemize}
}
\item{ Upgrades are underway at the Vancouver \textit{MOST} mission ground station to support the BRITE satellites and improve mission up- and downlink capabilities. This additional station will add 30 - 40 MB/day to the downlink capacity. }
\end{enumerate}
\noindent Problem 4: Radiation damage and thermal fluctuations. The first full-frame images downloaded from UBr and BAb showed signs of radiation damage more widespread than had been anticipated (see Fig.~\ref{fig:firstlight}). 
The radiation effects are correlated with satellite temperature, which varies between 40$^{\circ}$ and 7$^{\circ}$C for the two Austrian nanosatellites. The radiation problem posed a major threat to the lifetime and effectiveness of the BRITE mission, and we describe the solutions in detail in Sect.~\ref{sect:radiation}. 

\subsubsection{BLb}
\label{sect:blbcomm}

The first seven steps of the commissioning process (Fig.~\ref{fig:sfl_commission}) proceeded in roughly the same manner as BAb and UBr. Commissioning of BLb was significantly hampered due to the UHF interference with European ground stations. Commissioning of the new ST-16 star tracker aboard BLb was supported by the manufacturer, Sinclair. The performance is significantly better than the AA-MST, working well even in the SAA. After normal commissioning, new auto-download software was installed on BLb, increasing the downlink rate to 21 MB/day.

\subsubsection{BMb and BTr}
\label{sect:btrcomm}

Inspection of NORAD telemetry suggests that BMb never separated from the Dnepr launch vehicle. There is no way to activate it while it remains attached to the upper stage. If it does detach in the future, it will be released into a highly inclined, higher altitude orbit which would reduce its useful life due to higher fluxes of damaging radiation. BTr was released into orbit as planned and all 13 stages of commissioning were complete within only 8 days. The rapid progress was thanks to (a) the experience of the previous BRITE commissioning including the recognition of major software bugs, and (b) the higher downlink rates possible with the SFL ground station.


BMb and BTr were equipped with additional radiation shielding compared to earlier BRITEs (see Sect.~\ref{sect:radiation}) and this reduced the radiation-induced damage to BTr's CCD. This was evident very early its mission, as shown by a comparison of a sample BTr subraster to comparable subrasters from the other spacecraft (Fig.~\ref{fig:firstlight}).



\subsubsection{BHr}
\label{sect:bhrcomm}

Following a successful launch, communications were swiftly established during the first pass of the satellite over the ground-station in Warsaw. From the outset, transmissions in both the uplink and downlink directions were strong. BHr, having a slightly different telescope setup, required more adjustments to achieve a pointing accuracy consistent with mission standards but was completed roughly 3 months after launch.  BHr, which like BTr also has additional shielding, is also much less affected by radiation damage (Fig.~\ref{fig:firstlight}).

\section{Operations}
\label{sect:operations}

Day-to-day operations are handled by the teams appointed by the three partner nations in the BRITE mission, coordinated since April 2016 by TUG (Technical University Graz). The standard operations are the same for all BRITEs. Each primary field is observed
by at least one blue and one red filtered satellite in tandem. Depending on need, the remaining satellites will either observe other fields or complement observations of the primary field. The exact statistics of what has been observed to date is available on the Brite Photometry Wiki\footnote{\url{http://brite.craq-astro.ca/doku.php}}.


\subsection{Target Field Selection and Acquisition}

In 2007, the BRITE Executive Science Team (BEST)\footnote{The responsibilities of BEST are outlined in the BRITE Bylaws \url{http://www.brite-constellation.at})} invited BRITE proposals from the astronomical community. BEST reviewed and ranked the submitted proposals and selected targets and fields for the first year of the mission.  When Poland joined BRITE Constellation in 2010, another call for proposals was sent specifically to the Polish astronomical community. As of writing, there are 42 approved proposals with Principal Investigators (PIs) from 13 countries, which include more than 6,300 stars brighter than V = 6.  Given the luminosity selection effect of a sample brighter than the naked-eye limit, there are many O, B and Wolf-Rayet stars, and red giants, but also many other classes of stars are represented.

Observing fields are selected on the basis of the distribution on the sky of higher-priority science targets. Fields are selected a minimum of 1 year in advance to enable planning of supporting ground-based observations. From the vantage point of a BRITE satellite, a given field is visible for no more than 6 months at a time. The coordinate center of a chosen field and the instrument's boresight roll angle are fine-tuned to optimize the satellite fine pointing performance.

The fields monitored by the Austrian satellites, which feature the AeroAstro MST star tracker, must be close to the Galactic plane in order to include several very bright stars (V \textless 3). While the other BRITE satellites do not face this restriction -- see Tab.~\ref{tab:compare} for comparison of parameters to mission requirements-- this region of the sky contains the largest number of targets for the mission science goals, so many BRITE fields are in fact located in or near the Milky Way.


According to the Bylaws, BRITE observing fields are established at least 12 months beforehand, in order to facility planning of complementary observations. If a field is not selected well in advance, it is for one of two reasons:

\begin{itemize}
\item{The field contains a proposed Target of Opportunity and approved by BEST.}
\item{A planned field cannot be observed by a satellite for unanticipated technical reasons. If the problem persists, the satellite is reassigned to a new field or a previously observed one.  The redundancy of 5 satellites with 2 filters means the Primary Science Field is still monitored over a long time span, but with reduced cadence.}
\end{itemize}

After pointing a BRITE instrument at the central coordinates of a target field and setting the roll angle of the satellite, trials are conducted to confirm it has achieved fine pointing.  If it has not, then field center and roll angle offsets are applied until fine pointing lock is successful. This usually requires about a day, but for some satellites and fields (for reasons described earlier), it can take up to 2 weeks. Then a full-frame image is downloaded to confirm the field positioning is correct and all intended target stars are in the frame.  Subrasters are positioned around each target star for data download, as defined by a setup file uploaded to the IOBC.

\subsection{Data Acquistion}

The exposure time is selected to give good signals for stars in the field which span a wide dynamic range of brightness (down to V = 4, sometimes V = 7).  A 1-second exposure is best suited for the range 0 \textless V \textless 4; for fields with fainter targets and where the pointing is of sufficient stability, a 5-second exposure is used.  For fields with bright and faint targets, multiple exposure times are used, with the longer exposure time most often dominating the majority of the time within an observing window and a shorter taken at the beginning and/or end to obtain unsaturated bright star PSFs. This may change based on the the relative number of bright to faint star in a given observation field.  Multiple satellites monitoring the same field can be assigned different exposure times appropriate to their filters.

Pre-launch plans for BRITE included on-board processing of the subraster data, so that compressed photometry could be downloaded to efficiently  use the limited downlink capacity (2 - 4 MB/day).  The occurrence of warm pixels and evolving radiation damage on the BRITE CCDs meant that on-board photometric analysis was not feasible, and downloading resolved images was necessary to permit robust data reduction. Downloading 15 subrasters (32 $\times$ 32 pixels, 14.4$\arcmin$ $\times$ 14.4$\arcmin$ each), taken 3 times per minute for at least 15 minutes of each BRITE orbit, for 14 orbits per day, represents a data rate of about 10 MB/day. Upgrades to the European ground stations (see Section 3.2.1) and the higher capacity of the Toronto SFL station have allowed the mission to accommodate the higher downlink needs.


There have been attempts to reduce the downlink rate while maintaining data quality, such as stacking several consecutive exposures (adding them pixel by pixel) on board and downloading the stacks. Stacks of 3, 5 and 10 exposures were tested, and the stacked data were of poorer photometric quality (see Paper III).   Stacking is a `last resort' that was used only earlier in the mission when the European ground stations had lower downlink capability.


The hot pixels and pixel clusters (see Sect.~\ref{sect:cos_damage}) on the CCDs caused by radiation damage introduced scatter and systematic effects in the photometry of target star images.  To compensate for this, a very effective strategy has been adopted for BRITE measurements.  Rather than maintaining pointing on the same position in the field, the telescope alternates between two positions roughly 0.2 degrees apart. In consecutive images, the PSF of a target star shifts to different pixels, while pixel defects remain in the same positions.  Subtraction of these images effectively removes the effects of the defective pixels, as shown in Figure~\ref{fig:hurdle_chop}. To this aim, the target star subrasters are increased in size to 48 x 24 (21.6$\arcmin$, 10.8$\arcmin$) pixels, which uses more of the downlink bandwidth but with the reward of better photometry. This ``chopping" technique -- inspired by the approach used in ground-based infrared photometry -- is now used with all BRITE satellites and is discussed in greater detail in Section \ref{sect:chop}.

Along with the CCD image data, various spacecraft telemetry parameters are also downloaded with each measurement, including temperatures from 4 sensors on the CCD, local magnetic field strength, and position of the satellite in its orbit.  The onboard computer is able to store 10 days of continuous observations before data are overwritten, which allows for occasional backlogs due to ground station outages, etc. All BRITE data are stored in an archive at the Copernicus Astronomical Center in Warsaw, Poland. The data are reduced by the BRITE team and provided to the PIs by BEST. The format of the reduced data files, and the pipeline used to generate them, are described in detail in Paper III.

\begin{table}[!h]
\centering
\caption{Current status of operations versus original mission requirements.}\label{tab:compare}
\vspace{3mm}
\begin{tabular}{ccc}
\hline\noalign{\smallskip}
Parameter & Goal & Achieved\\
\noalign{\smallskip}\hline\noalign{\smallskip}
Download rate & 10\,MB/day & $>$\,40\,MB/day\\
FTAP stability & 1.5\,$\arcmin$ (3\,pixels, 1$\sigma$) & $<$\,12$\arcsec$\\
Subraster & 32\,$\times$\,32\,pixels & 24\,$\times$\,24\,pixels \\
\hspace{4pt} chopping & & 48\, $\times$\,24\,pixels\\
Magnitude range & 0\,--\,4 & 0\,--\,7\\
Sky accessibility & all-sky & minor constraints\\
Coverage per orbit & 15\,min & 30\,min\\
Targets per field & 15 & 15--37 \\
Fields per orbit & 1 & 1--2 \\
\noalign{\smallskip}\hline
\end{tabular}
\end{table}

\subsection{Supporting ground-based observations}
\label{sect:gbot}

The BRITE Ground-Based Observing Team (GBOT)\footnote{More information can be found at \url{http://www.brite-constellation.at}} organises and coordinates observing runs to support the on-orbit photometry. The supporting data include spectroscopy, spectropolarimetry and photometry with other bandpasses.  Currently, GBOT has 51 members from 15 countries.  One example of a GBOT project which supports many BRITE targets is the {\it BRITE Spectropolarimetric Survey} conducted at the Bernard Lyot Telescope, the ESO 3.6-m telescope, and the Canada-France-Hawaii Telescope (CFHT). Started in 2014 and scheduled to be completed in 2016, the goal of this survey is to obtain a high-S/N, high-resolution circular polarisation spectrum of every star in the sky brighter than V = 4 \citep{neiner14}.  The data will be used to detect new magnetic fields in BRITE targets and derive their strength and configuration.

\section{CCD radiation damage: Problems and solutions}
\label{sect:radiation}

Satellites operating in low Earth orbit are exposed to energetic protons and electrons trapped by the Earth's magnetosphere. This particle radiation field can cause damage to electronics, memory and CCD detectors, which accumulates with time \citep[see][report on the Hubble Space Telescope]{hubb_rad}.  Noise in a CCD is caused by ionizing particles (``cosmic rays") which generate false signals when they deposit their energy as charges in the CCD pixels. Displacement damage occurs when a cosmic ray displaces a silicon atom from its lattice in the CCD substrate, creating a persistent defect \citep[see][for a comprehensive review]{rad_rev}. This can introduce what are known as ``hot pixels" and ``warm columns", and it can increase charge transfer inefficiency (CTI).


The small size and mass of a BRITE nanosatellite limited the amount of radiation shielding of the CCD possible, and the low mission budget meant that extensive radiation testing could not be done before launch. However, it was expected that the Kodak CCDs used were radiation tolerant.  The first downloaded images (see Fig.~\ref{fig:firstlight}) were a surprise, showing much more radiation damage than anticipated in a short time. Several combined solutions to address this problem are now in place. Those that are applied on board the satellites are described below.  Those added to the data reduction pipeline on the ground are discussed in Paper III.

\subsection{CCD cosmetic damage}
\label{sect:cos_damage}	
\label{sect:hps}

Displacement damage in the region of the CCD exposed to light creates gaps which more easily generate electrons thermally, without any photon input, resulting in pixels with false signals several standard deviations above the background. The signal $D$ from thermal electrons increases exponentially with CCD temperature, according to:
\begin{equation}
\label{eq:weird}
D = Gt_{\rm e}\exp(-E_{\rm a}/kT),
\end{equation}
where $t_{\rm e}$ is the exposure time, $E_{\rm a}$ is the activation energy, $T$ is the temperature of the detector (in Kelvin),  $k$ is the Boltzmann constant and $G$ is a constant \citep{janesick01}. 
Hot pixels can sometimes exhibit Random Telegraph Signals (RTS) where $G$ oscillates among a set of discrete values, so the signal from an RTS pixel cannot be predicted \citep{RTS, RTS2}. Thermal electrons in the charge register portion of the pixel result in increased dark current when charges are transferred from pixel to pixel. This creates a ``warm column" on the CCD whose background is higher than neighboring columns.


\begin{figure}[htbp!]
\centering
\includegraphics[trim=1cm 0cm 2.5cm 0cm,clip,width=0.45\textwidth]{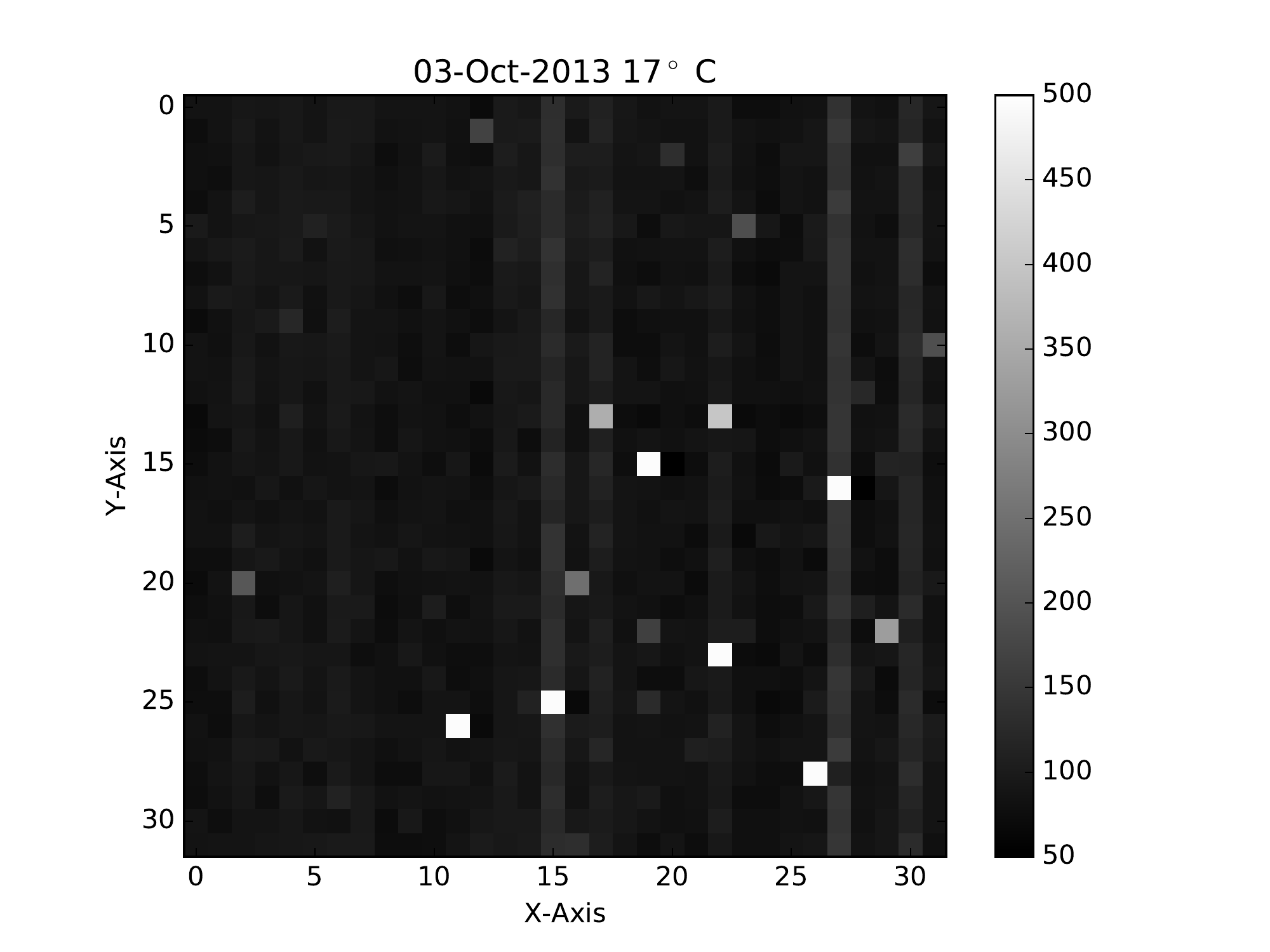}\label{fig:hp_lowT}
\includegraphics[trim=1cm 0cm 2.5cm 0cm,clip,width=0.45\textwidth]{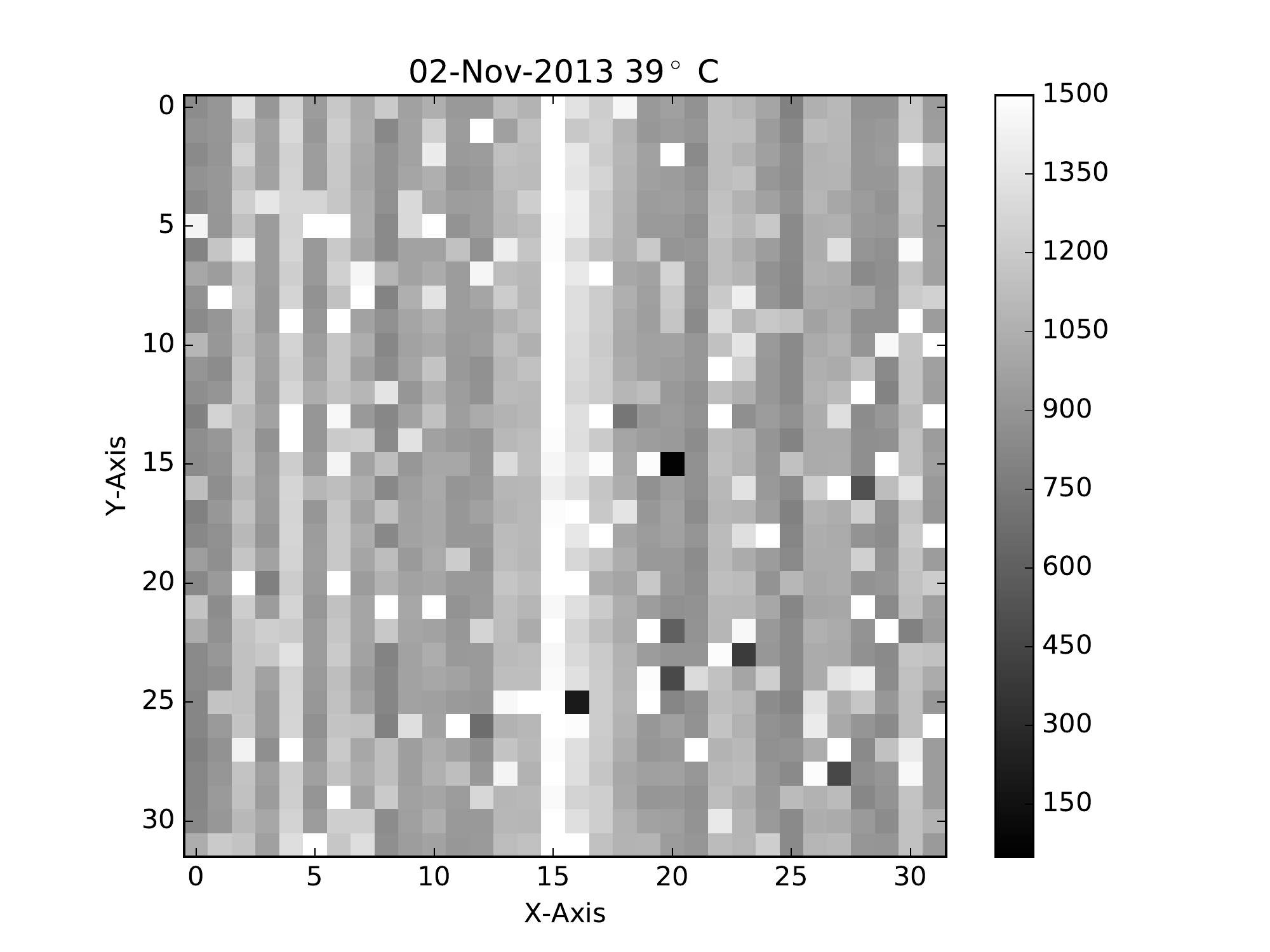}\label{fig:hp_hiT}
\caption{\small{Dark frames from the UBr CCD early in the mission, specifically the subraster of target HD 36486 in Orion. At left is an exposure at CCD temperature 17$^\circ$C; at right, an exposure of equal length at 39$^\circ$C. Even at the lower temperature, there are more than a dozen hot pixels and at least 5 warm columns visible to the eye. An increase in temperature of 22$^\circ$C results in a tremendous increase in the number of defects. Because the background increases significantly with temperature, the contrast has been lowered by a factor of 3 in the left image to make radiation effects visible.}}
\label{fig:hurdle_hp}
\end{figure}
With sufficient shielding, and if the CCD is operated at a sufficiently low temperature, such defects should not accumulate quickly for a detector in low Earth orbit.  In many cases, the defects are not permanent, and can be removed by warming the CCD in a process called ``annealing" \citep[see][]{hubb_rad}. However, the BRITE detectors can have only modest shielding, there is no active cooling of the CCDs, and the trim heaters can bring the CCDs to a peak temperature of only 60$^\circ$C, insufficient for any effective annealing of cosmetic damage.  Figure 10 shows the level of radiation damage on a BRITE CCD and the sensitivity of the warm pixel noise to CCD temperature.  For the first BRITE nanosatellites launched, BAb and UBr, the fraction of the active CCD area affected by warm pixels increases by 0.0042\% per day. At this rate, 10\% of the pixels will be defective in about 7 years.


\subsubsection{Increased Charge Transfer Inefficiency (CTI)}
\label{sect:cti}

\begin{figure}[htbp!]
\centering
\includegraphics[trim= 1cm 2cm 2cm 0cm,clip,width=0.45\textwidth]{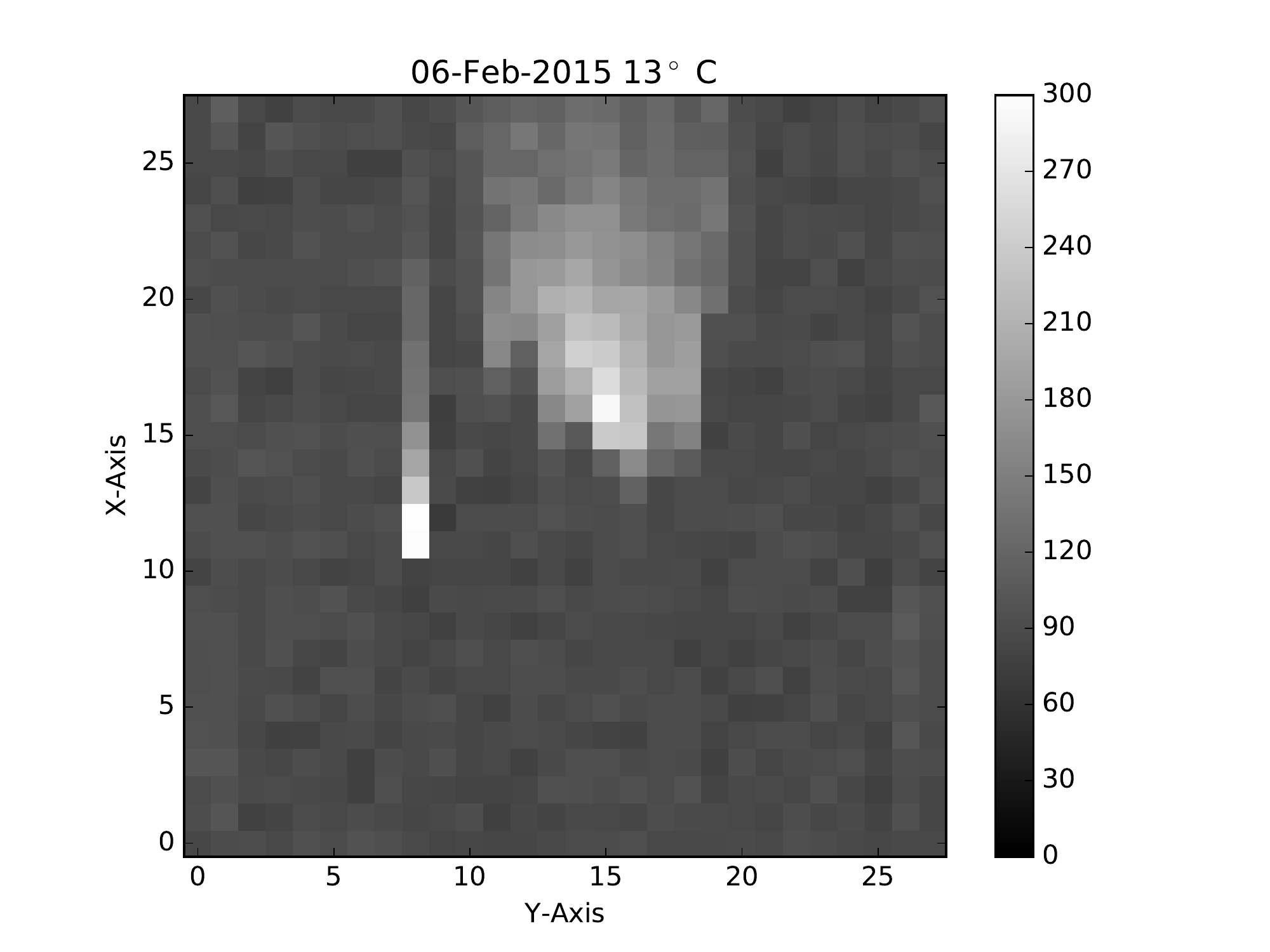}\label{fig:cti2_im0}
\includegraphics[trim=1cm 2cm 2cm 0cm,clip,width=0.45\textwidth]{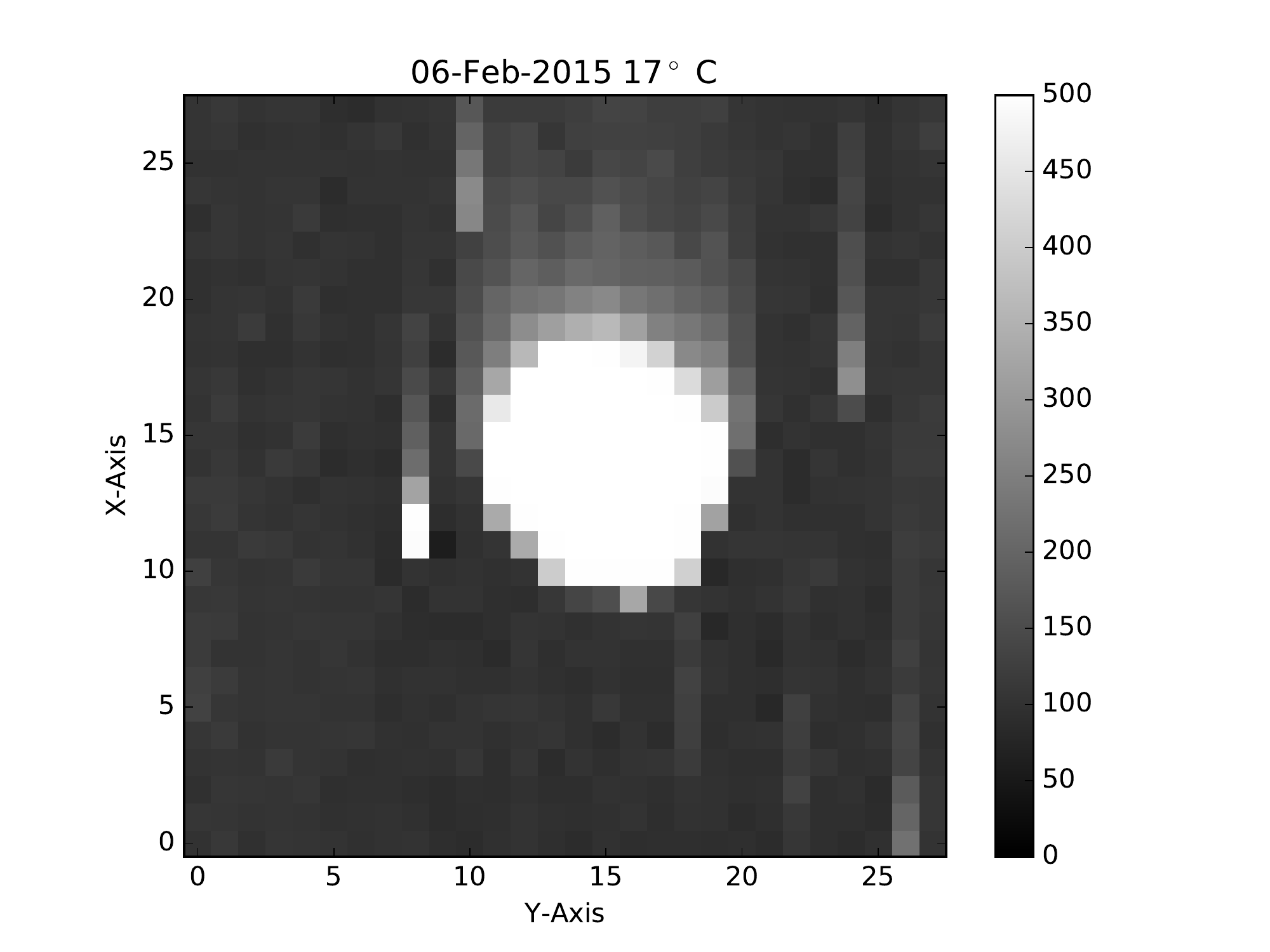}\label{fig:cti2_im62}
\caption{\small{BTr images of Cygnus target HD 193237 with the same exposure time but at temperatures of 13$^{\circ}$C (left) and 17$^{\circ}$C (right).  The decrease in CTI is evident as the temperature increases showing significantly less blurring in the image at 17$^{\circ}$C. Moreover, the drop in signal is so pronounced in the subraster at 13$^{\circ}$C that the scale has to be reduced from 500 to 300 ADU to make the PSF visible.}}
\label{fig:hurdle_cti}
\end{figure}

If cosmic ray penetration is shallow enough that displacement damage is confined to the substrate surface, photoelectrons are impeded from moving from pixel to pixel during readout.  The defects, known as ``traps", introduce different delays in the transfer of the trapped electrons, producing streaks along columns during readout.  All CCDs have a very small inherent Charge Transfer Inefficiency (CTI), but a much larger than expected CTI was seen in the images from BTr only two months after its launch (see Fig.~\ref{fig:hurdle_cti}).  Careful examination of images from the other BRITE satellites also revealed CTI that was higher than specified by the manufacturer. After little more than a year in orbit, roughly $\frac{1}{4}$ of the UBr detector was affected by noticeable levels of CTI. Lower CCD temperature means less mobility of electrons, so CTI becomes less severe on a warmer CCD (see Fig.~\ref{fig:hurdle_cti}).  Hot pixels and warm columns conversely, worsen on a warmer CCD.  


\subsection{Mitigating radiation damage}
\subsubsection{Controlling temperature without a cooling system}
\label{temp-stability}

\begin{figure}[htbp!]
\centering
\includegraphics[trim=0cm 0cm 0cm 0cm,clip,width=1.0\textwidth]{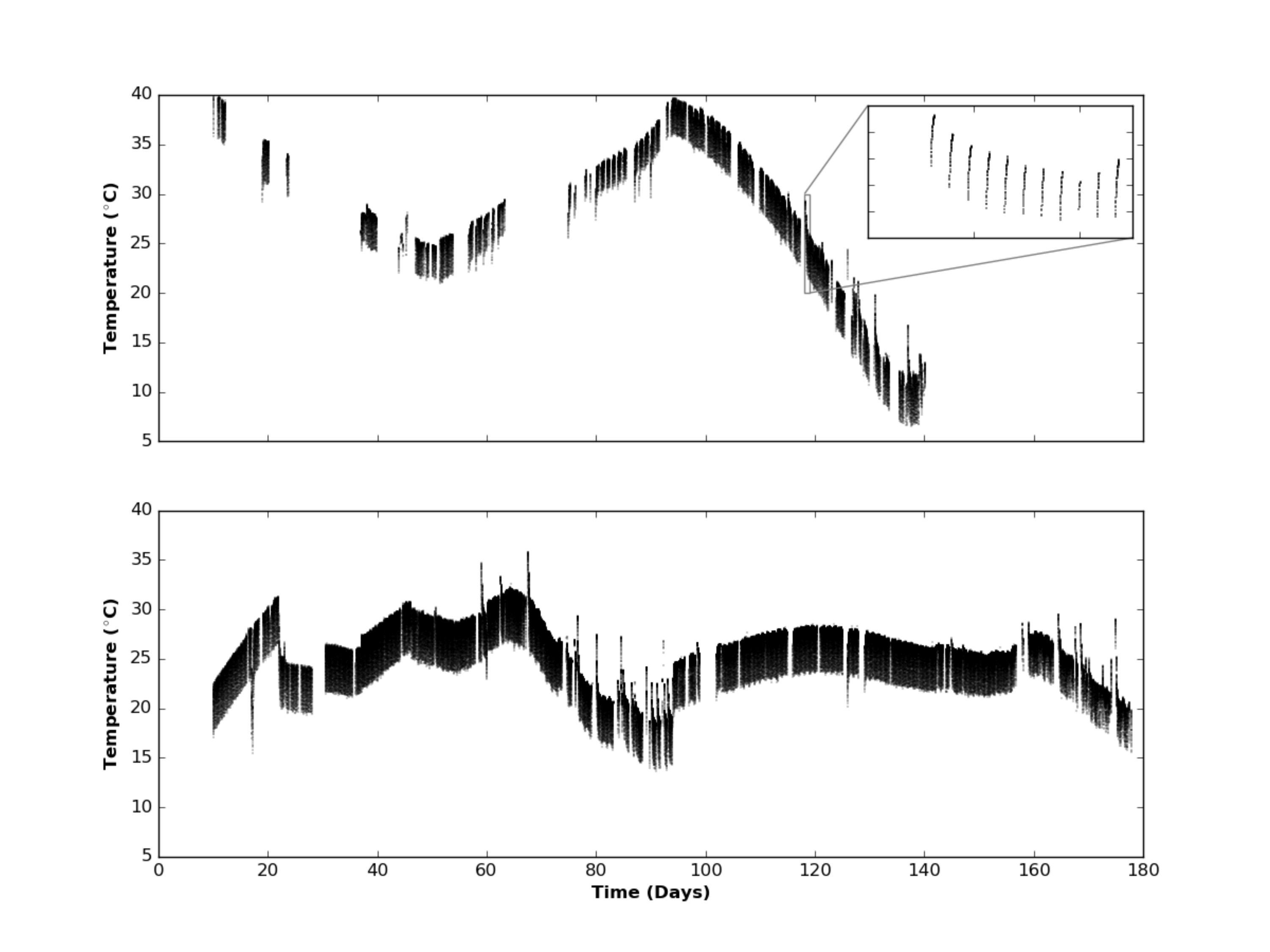}
\caption{\small{CCD temperatures for the UBr satellite before (top) and after (bottom) the anti-Sun pointing strategy was adopted.  The former data were collected during monitoring of the Orion I field and the latter during Perseus field observations. As seen in the zoomed inset of top figure, temperature modulations during each satellite orbit account for the  widths of the two curves. The plotted data are the averages of temperatures from the 4 sensors on the CCD.}}
\label{fig:temp_fluc}
\end{figure}

The temperature swings during a BRITE satellite orbit translate into drastic changes in the levels of cosmetic defects and CTI.  This is most pronounced for the satellites whose orbits have the longest periods of uninterrupted sunlight (BAb, UBr and BLb). Without any active cooling mechanism, the only way to reduce these temperature variations is to reorient the spacecraft with respect to the Sun.  At the end of each observing segment in a BRITE orbit, the telescope is pointed directly away from the Sun, so that a minimal area of the spacecraft is exposed to sunlight. The effects of this strategy can be seen in Fig.~\ref{fig:temp_fluc}.  The mean CCD temperature during one orbit is roughly the same, but the range of temperature variation is reduced by half.


\subsubsection{Increased shielding}

The commissioning of the Austrian BRITE satellites identified a  number of problems which could be addressed by changes in the BRITE satellites which had not yet flown.  Modifying the accepted designs, and changing hardware which had already been built, can never be done lightly for any satellite.  However, the BRITE satellite designs could accommodate a few limited changes to increase shielding near the CCD. 

To confirm that the damage was due to high-energy protons, and to evaluate how effective new shielding would be, radiation tests were conducted at the TRIUMF cyclotron facility on the campus of the University of British Columbia (Vancouver, Canada). An unpowered BRITE flight-spare CCD was exposed to 0.5 krad of 100-MeV protons. 
The TRIUMF test confirmed that protons were the culprit, and the effectiveness of different materials and  thicknesses of shielding against protons were tested.

For the unlaunched Canadian and Polish BRITE satellites, different plans were enacted. Final qualification testing of the Canadian nanosatellites had been completed, so there were few options for redesign.  The best option was to replace the aluminium CCD header tray with an identical tray made of tungsten, which would reduce the proton flux reaching the detector by a factor of 2.75.

When BTr was launched and began operating, it was soon clear that the tungsten shielding had the desired effect, reducing significantly the rate of hot pixel generation. However, the shielding intercepts only particles incoming at high impact angles relative to the CCD plane.  It offers little protection against particles arriving at shallow impact angles, so CTI is still a concern.  Indeed, increased CTI was measured on the BTr detector after a few months in orbit.



At this point in the mission, BLb had already been launched, so no changes in the shielding were possible. For BHr, which has a different optical design with a more compact telescope in the same nanosatellite volume, there was room to add additional shielding. Tungsten is effective but it is dense, so introducing a thick tungsten shield to block shallow particle hits would have increased the payload mass by too much. Tests carried out at the National Nuclear Research Office (Poland) showed that a less dense material -- hydrogen-rich polyethylene made of Borotron -- would be effective for BHr shielding and was implemented  (Mochnacki et al. 2016, in prep.).


\subsubsection{Chopping}
\label{sect:chop}

It was possible to add increased shielding on only 3 of 5 operating BRITE satellites, so other non-hardware strategies were necessary to mitigate the effects of CCD radiation damage on photometry. A dark frame after each exposure, subtracted from that exposure, would correct for the background (including cosmetic defects). However, this would require the satellite to leave fine pointing and then reacquire the target field after the dark frame. This process would consume several minutes and it would significantly reduce the number of measurements made per orbit. 

\begin{figure*}[!htbp]
\centering
\includegraphics[trim=2cm 0cm 2cm 0cm,clip, height=0.25\textheight, width=0.30\textwidth]{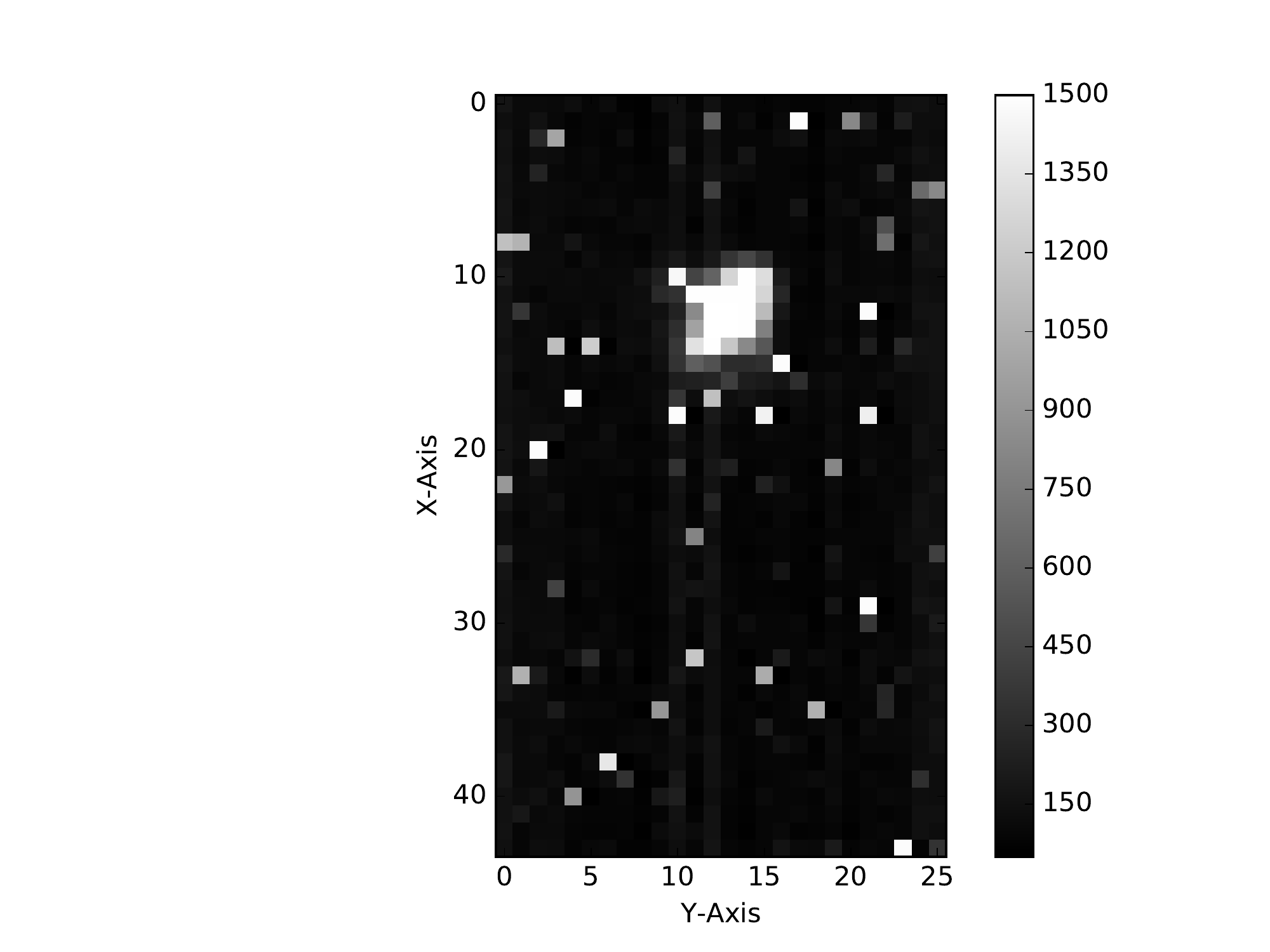}\label{fig:stari}
\includegraphics[trim=2cm 0cm 2cm 0cm,clip, height=0.25\textheight, width=0.30\textwidth]{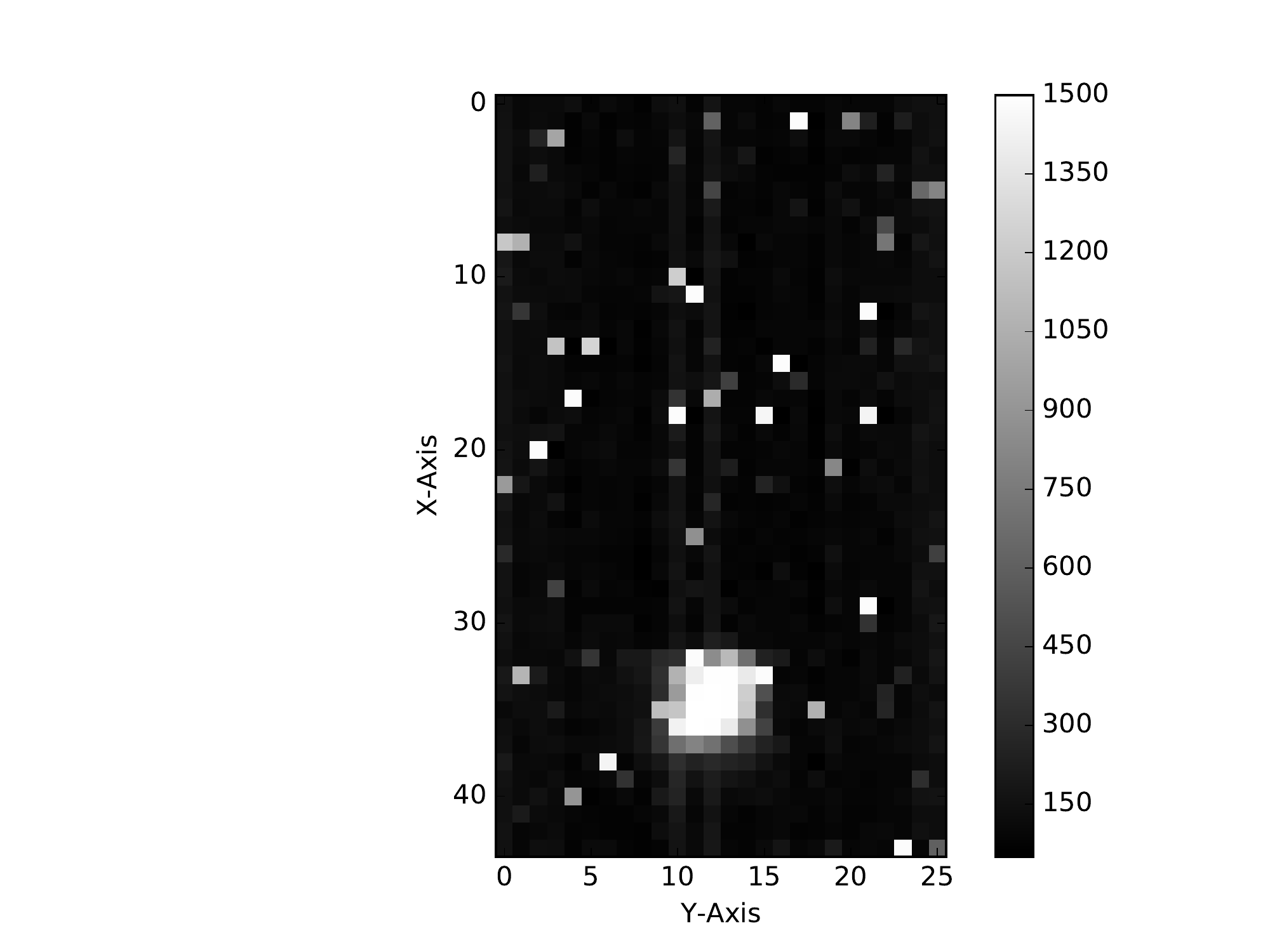}\label{fig:starf}
\includegraphics[trim=2cm 0cm 2cm 0cm,clip, height=0.25\textheight, width=0.30\textwidth] {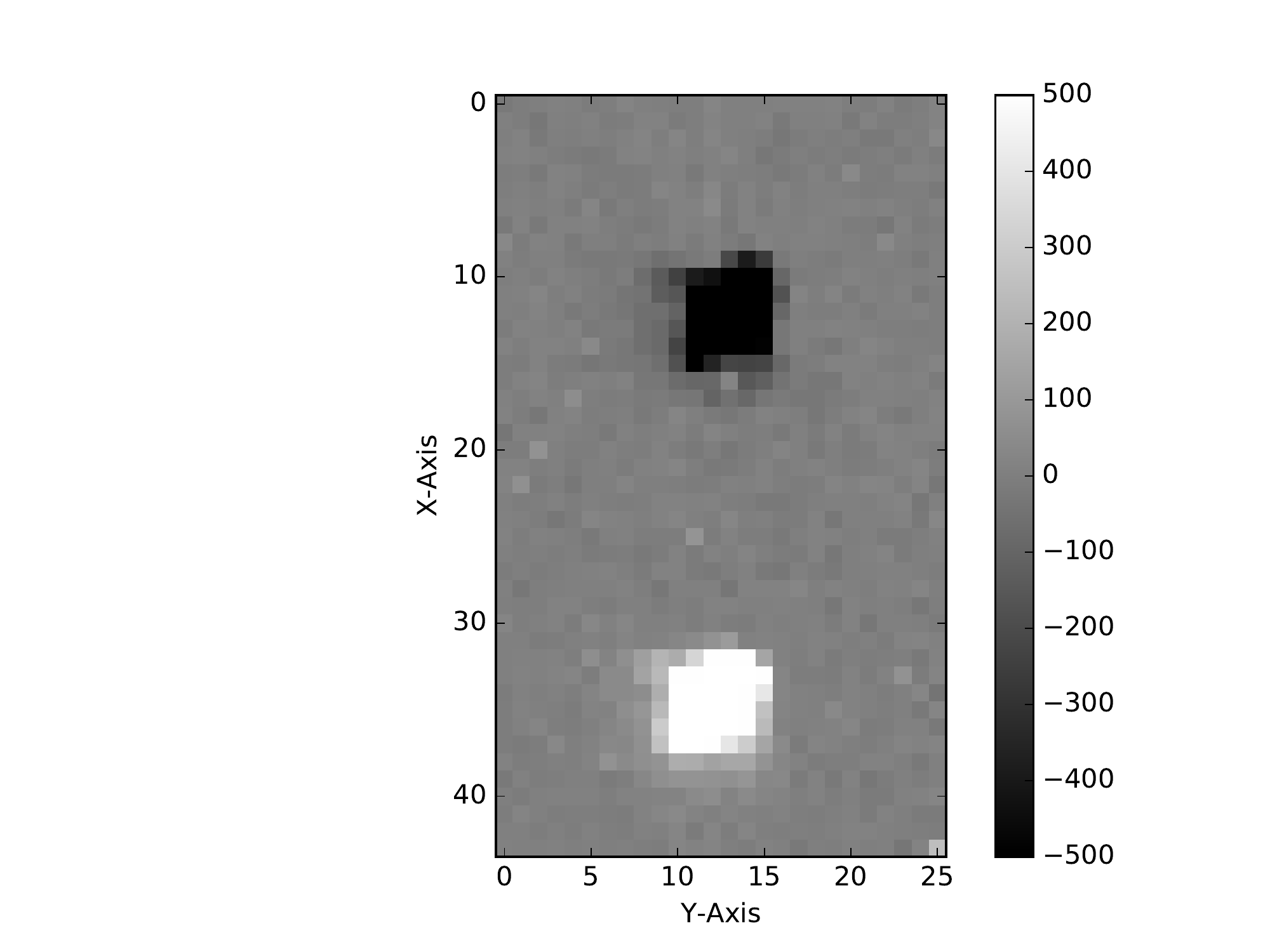}\label{fig:chsub}
\caption{\small{The power of the chopping technique for BRITE photometry.  (left and middle) Subrasters of consecutive exposures of the target HD 23630, where the telescope pointing has been deliberately offset.  The star PSF dominates these frames; the other lit pixels are radiation-induced defects. (right) Subtraction of left from middle. The greyscale of the difference frame has been changed to show clearly the positive and negative PSFs of the star, and the near-complete filtering of CCD defects.}}
\label{fig:hurdle_chop}
\end{figure*}

A more time-effective solution is chopping, inspired by ground-based infrared imaging astronomy, where the thermal background of the telescope itself is a serious factor. While this is accomplished in infrared astronomy by movement of the secondary mirror, BRITE achieves a similar result through offsetting the telescope pointing back and forth in successive exposures. For BRITE, the target star is kept in the subraster in both offsets but the PSFs do not overlap (left and middle panels of Fig.~\ref{fig:hurdle_chop}). When the two offset frames are subtracted, the result is one image with one positive and one negative target PSF, where the background defects have been mostly removed (see right panel of Fig.~\ref{fig:hurdle_chop}). 

There are costs to chopping.  (1) The center of the field is adjusted slightly between each exposure, increasing the minimum time between measurements.  Fortunately, this minimum time is only about 20 seconds, comparable to the enforced idle time already in place due to ground station downlink limits.  (2) The subraster size is effectively doubled to accommodate the offset.  Again, with downlink limits, this means one can download fewer exposures of the same number of targets, or monitor fewer targets.   The improvement in data quality justifies this cost, and when the MOST UBC ground station in Vancouver joins the BRITE network, the downlink capacity will go up significantly and compensate for this drawback.  Chopping is even effective for RTS effects, since the typical lifetime of a pixel in a given RTS state is several minutes (Janesick 2001), much longer than the time between successive chopping exposures on any of the BRITE satellites.

\begin{figure*}[!htbp]
\centering
\label{fig:cti_sol}
\includegraphics[trim=0cm 0cm 0cm 0cm,clip,width=0.48\textwidth]{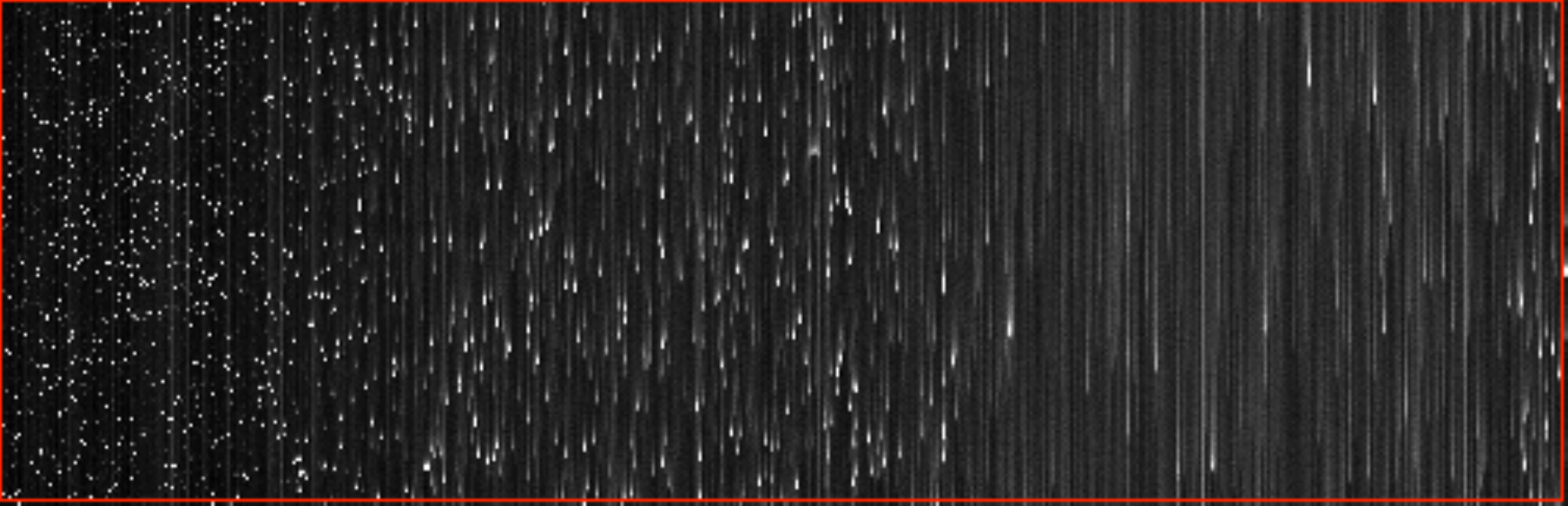}
\includegraphics[trim=0cm 0cm 0cm 0cm,clip,width=0.48\textwidth]{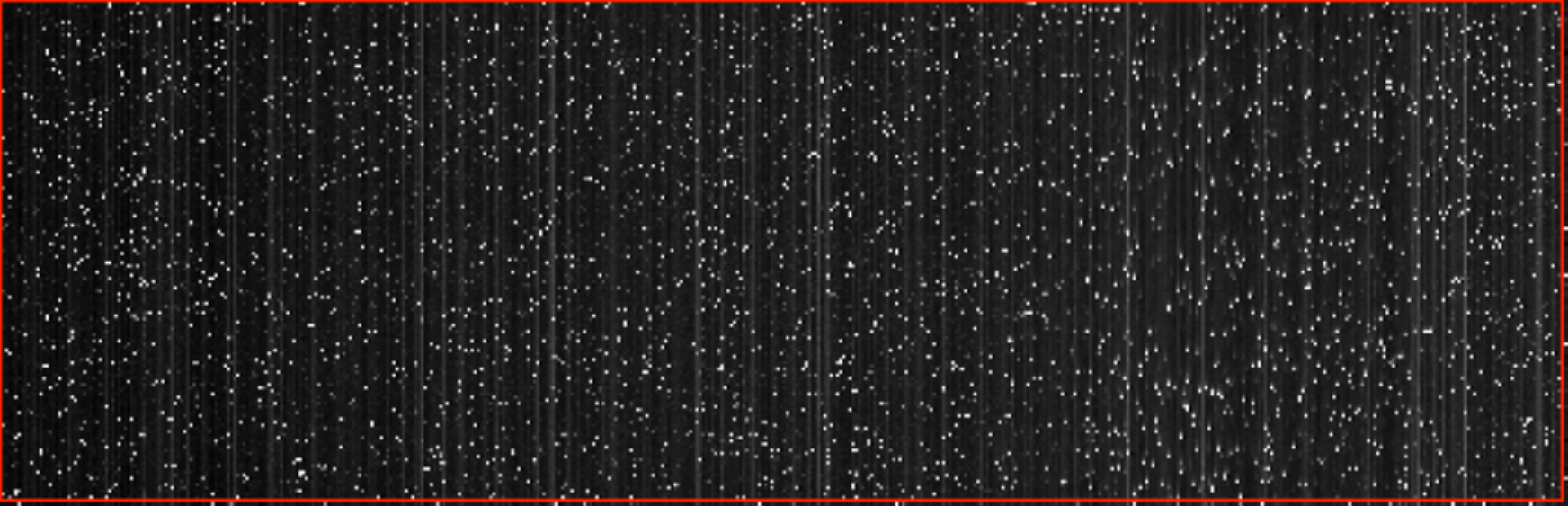}
\caption{\small{Subrasters of test images from UBr at CCD clocking times of 3.5 $\mu$s (left) and 20 $\mu$s (right), demonstrating the reduction of CTI at slower clocking rates.}}
\end{figure*}

Chopping does not correct fully for CTI effects, but there is a simple way to reduce the impact of CTI: slowing the readout time (CCD clocking time). This solution - tested by SFL before being implemented on the operational satellites - gives photo-electrons more time to navigate the traps near the substrate surface, increasing the transfer efficiency.
The readout time of each BRITE detector was increased from 3.5 $\mu$s to 20 $\mu$s (the maximum recommended by the manufacturer), and this change has effectively solved the CTI problem for BRITE (see Fig.~\ref{fig:cti_sol}).

\section{Summary and current mission status}

BRITE Constellation marks the entry of nanosatellite technology into space astronomy, with the lowest cost per payload for a space astronomy mission to date.  In addition to the scientific returns of BRITE, we see it (and this paper) as a model for future low-cost space science missions with nanosats. 

The lessons learned include the problems encountered on orbit and the strategies adopted to deal with them, as described in this paper. But one fundamental strategy was part of the original BRITE Constellation concept: multiple satellites with identical and complementary capabilities, offering redundancy in the event of the failure of one or more satellites.  This redundancy has been demonstrated clearly with the success of BRITE despite the loss of one of six satellites in the Constellation.  Lower cost generally means higher risk, but the experiences of BRITE, and MOST before it, show how risks can be mitigated even on a low budget and a rapid schedule for a space satellite mission.

As this paper goes to press, all five operating BRITE satellites are meeting (and in some ways, exceeding) the original mission requirements. Initially severe problems due to CCD radiation damage have been much alleviated, thanks especially to chopping and increasing clocking time. (Additional details on how BRITE meets its photometric requirements through data reduction are given in Paper III.) As of March 2016, observations have been completed on 11 fields with 2 fields monitored twice, covering more than 300 stars.  All but one field\footnote{Observation field 3, Sagittarius, was only observed for 42 days.} were observed in two colours, with time spans from 108 to 178 days. Two fields are currently being observed by BRITE Constellation and observations are currently planned through the summer of 2017.



Even if the mission were to end today, BRITE Constellation is already a success.  All indications are that the mission can continue well beyond the originally budgeted 2-year lifetime, if funding continues.  The future of BRITE is bright, and thanks to BRITE, so is the future of low-cost space-based astrophysics.


\begin{acknowledgements} 
AFJM is grateful for financial aid from CSA, NSERC and FQRNT (Quebec).OK, RK and WW are grateful for funding via the Austrian Space Application Programme (ASAP) of the Austrian Research Promotion Agency (FFG) and BMVIT; and WW for support of the University of Vienna (IS 538001, IP 538007). KZ acknowledges support by the Austrian Fonds zur Foerderung der wissenschaftlichen Forschung (FWF, project V431-NBL). The research grant support to SMR from NSERC of Canada was used to fund testing and implementation of the optimum CCD pixel readout time. The Polish co-authors gratefully acknowledge support from the BRITE PMN grant 2011/01/M/ST9/05914; GH additionally thanks the NCN grant 2015/18/A/ST9/00578. JMM and GAW acknowledges Discovery Grant support from NSERC. AP acknowledges the NCN grant No. 2011/03/B/ST9/02667. APo was supported by the Polish National Science Center, grant no. 2013/11/N/ST6/03051. 
\end{acknowledgements}
\bibliography{biblio} 

\end{document}